\documentclass[aps,twocolumn,pra,showpacs,amsmath,amssymb,longbibliography,showkeys,10pt]{revtex4}
\usepackage{graphicx}
\usepackage{epstopdf}
\usepackage[normalem]{ulem}
\usepackage{braket}
\usepackage{hyperref}
\usepackage[dvipsnames]{xcolor}
\usepackage{ulem}
\allowdisplaybreaks

\begin{document}

\title{Quantum jumps in amplitude bistability: Tracking a coherent and invertible state localization}

\author{Th. K. Mavrogordatos}
\email{themis.mavrogordatos@fysik.su.se;\\ th.mavrogordatos@gmail.com}
 \affiliation{Department of Physics, AlbaNova University Center, SE 106 91, Stockholm, Sweden}

\date{\today}

\begin{abstract}
We investigate the nature of quantum jumps occurring between macroscopic metastable states of light in the open driven Jaynes--Cummings model. We find that, in the limit of zero spontaneous emission considered in [H. J. Carmichael, Phys. Rev. X {\bf 5}, 031028 (2015)], the jumps from a high-photon state to the vacuum state entail two stages. The first part is coherent and modelled by the localization of a state superposition, in the example of a null-measurement record predicted by quantum trajectory theory. The underlying evolution is mediated by an unstable state (which often splits to a complex of states), identified by the conditioned density matrix and the corresponding {\it quasi}probability distribution of the cavity field. The unstable state subsequently decays to the vacuum to complete the jump. Coherence in the localization allows for inverting the null-measurement photon average about its initial value, to account for the full switch which typically lasts a small fraction of the average cavity lifetime; an asymptotic law for the jump time is established in high-amplitude bistability. This mechanism is contrasted to the jumps leading from the vacuum to the high-photon state in the bistable signal. Spontaneous emission degrades coherence in the localization, and prolongs the jumps.
\end{abstract}


\maketitle

\section{Introduction}

Quantum jumps were conceived by Niels Bohr in 1913~\cite{Bohr1913}, at a time when experimenters could envisage light sources emanating {\it quasi}monochromatic blackbody radiation at best~\cite{Carmichael2001}. In that background was formulated the Einstein A and B theory~\cite{Einstein1916, Einstein1917}, together with the bold proposal by Bohr, Kramers and Slater that the wave amplitude was to determine probabilities for discrete transitions---quantum jumps---between stationary states~\cite{BKS1924I, BKS1924II}, and, ultimately, Schr\"{o}dinger's objection to the idea of an abrupt transition~\cite{Schrodinger1952}. With the advent of lasers, quantum jumps in trapped ion experiments involved scattering between atomic states, directly monitored by a ``telegraph signal''~\cite{Cook1985,Javanainen1986,Javanainen1986II, Nagourney1986, Sauter1986, Bergquist1986, CohenTannoudji1986, Cook1988, Itano2015}. Quantum jumps have since been observed in several atomic ~\cite{Finn1986, Reynaud1988, Erber1989, Carmichael1989, Basche1995, Peil1999, Gleyzes2007, Guerlin2007} and solid-state configurations~\cite{Jelezko2002, Neumann2010, Robledo2011, Hatridge2013, Sun2014}. Photons interacting with a readout cavity in the experiment of~\cite{Vijay2011} populated a reflected signal carrying information on a continuously observed qubit state in real time.

In parallel with the above line of research, the study of optical bistability was equally strongly tied to the development of quantum electrodynamics~\cite{Narducci78,Bonifacio1978,Lugiato1984,Dykman1979, Dykman1988,CarmichaelBook2}. For a light-matter coupling strength exceeding photon loss by two orders of magnitude, dwell times in macroscopic metastable states~\cite{Dykman2012, Peano2014} of several seconds can be attained~\cite{Sett2024}. In this picture of enhanced coherence, where do switching events between those states fit and what determines and their timescale? How do they correlate with individual quantum events like registered photon ``clicks''~\cite{Glauber1963,Glauber1963II,Kimble1977,YurkeStoler1986,Carmichael1993QTI,Carmichael1993QTII,Carmichael1999,Denisov2002,Foster2003,
CarmichaelBook2, CarmichaelBook1,Walls1981,Carmichael1985,Hanschke2020}? 

The results reported from the circuit QED experiment conducted by Minev and collaborators~\cite{Minev2019} for a case that corresponds to the original observation of quantum jumps in atomic physics~\cite{Bohr1913, Dehmelt1975} ``demonstrated that the evolution of each completed jump [between two states of an artificial atom] is continuous, coherent and deterministic''. Further, the authors ``exploited these features, using real-time monitoring and feedback, to catch and reverse quantum jumps mid-flight---thus deterministically preventing their completion''. In this configuration, quantum jumps between two levels were monitored by an auxiliary state. Drawing an analogue with quantum bistability~\cite{Armen2009,Sett2024}, we will find that switching between the high-photon metastable state to the vacuum is mediated by a third state, unstable to fluctuations yet participating in a coherent superposition. How can the sense of ``catching a jump mid-flight'' in~\cite{Minev2019} be interpreted for the case of bistable switching, and what possibilities are opened for measurement on the basis of coherence~\cite{Dykman2009}? 

In the present report, we aim to answer the above question by considering the mechanism underlying quantum jumps between macroscopic metastable states {\it dynamically} formed as a result of the inherent nonlinearity of light-matter interaction. The ``mid-flight'' attribute will acquire anew an operational meaning, on the basis of a null-measurement evolution~\cite{Dicke1981, Porrati1987, Mabuchi1996, Carmichael1993QTIV, CarmichaelBook2} which proves to be invertible. Before we speak of quantum-fluctuation switching along a single simulated or experimental run, in Sec.~\ref{sec:ME} we very briefly review the evolution of the system density matrix and the Maxwell--Bloch equations (MBEs)~\cite{Alsing1991, CarmichaelBook2, Carmichael2015}. In the limit of zero spontaneous emission, defining the ``zero system size'', the MBEs---also called neoclassical equations---identify new stationary states for resonant light-matter coupling with respect to absorptive bimodality, that play a key role in organizing quantum dynamics and the associated symmetry breaking~\cite{Alsing1991,CarmichaelBook2,Mabuchi2009,Curtis2021}. In Sec.~\ref{sec:cohlocinv}, we highlight the role of coherent localization in a quantum jump of Jaynes--Cummings (JC) bistability in the strong coupling limit. To familiarize ourselves with the operational consequences of a null-measurement record, in Sec.~\ref{subsec:nullempty} we set the light-matter coupling strength and external drive to zero, and consider instead an empty cavity mode prepared in a coherent-state superposition, ultimately damped to the vacuum. 

Our central results pertaining to the conditioned display of coherence in the JC bistability are presented in Sec.~\ref{subsec:cohlocJC}, ahead of making a detour to examine the regime of photon blockade as a forerunner of the distinction between upward and downward jumps. In the former (latter) the conditioned photon number in the cavity mode increases (decreases) as a result of a photoelectron ``click'' at the detector situated at the output field. We move on to examine how the conditional coherence manifests itself in a weak-coupling limit, focussing on the characteristic example of the Kerr nonlinearity in Sec.~\ref{sec:Kerr}. Photon correlations read by single realizations are examined in more detail in Sec.~\ref{sec:photonstat}, while the complementary unraveling scheme of heterodyne detection is given special emphasis. Common experimental limitations are taken into account in Sec.~\ref{subsec:sourcesincM}, before a setup of operationally extracting a conditional $Q$ function is discussed in Sec.~\ref{subsec:MQ}. It relies on averaging heterodyne detection records obtained from an empty cavity which is prepared in a coherent-superposition state during a downward jump of JC bistability. This method of record making allows for a direct connection to projective measurements, as shown in Sec.~\ref{subsec:proj}. The role of spontaneous emission is briefly discussed in the last section, before concluding remarks close the paper out.    

\section{Master equation and amplitude bistability in the strong-coupling limit} 
\label{sec:ME}

The coherently driven dissipative JC model~\cite{JaynesCummings1963, Paul1963, Alsing1991, Dutra1994, Carmichael2015, Larson24, Invariant2024, Burgarth2024} for ``zero system size'' is defined by the master equation (ME)
\begin{equation}\label{eq:ME}
\frac{d\rho}{dt}\equiv \mathcal{L}_{\kappa}^{\rm JCD}\rho=\frac{1}{i\hbar}[H_{\rm JCD},\rho] + \kappa(2a\rho a^{\dagger}-a^{\dagger}a\rho-\rho a^{\dagger}a),
\end{equation}
where $\rho$ is the reduced density operator for the system comprising a cavity mode resonantly coupled to a two-level system, and 
\begin{equation}
H_{\rm JCD}=-\hbar\Delta \omega (a^{\dagger}a + \sigma_{+}\sigma_{-})+i\hbar g(a^{\dagger}\sigma_{-}-a\sigma_{+})+i\hbar \varepsilon(a^{\dagger}-a)
\end{equation}
is the JC Hamiltonian in a frame rotating with the drive (see Appendix~\ref{sec:MC}); $a^{\dagger}$ and $a$ are creation and annihilation operators for the cavity mode, $\sigma_{+}$ and $\sigma_{-}$ are raising and lowering operators for the two-state system, and $g$ is the dipole coupling strength. Photons escape the cavity at an average rate of $2\kappa$. The drive field has amplitude $\varepsilon$, and its frequency $\omega_d$ is detuned with respect to the cavity mode by $\Delta\omega \equiv \omega_d-\omega_0$, where $\omega_0$ is the common frequency of the field mode and the two-state system. In this work, we operate under strong-coupling conditions, $g/\kappa \gg 1$, allowing for a well-defined JC nonlinear spectrum against the broadening induced by dissipation~\cite{Alsing1992, Tian1992, CarmichaelBook2, Fink2008, Bishop2009, Shamailov2010}. 

Bimodality in the steady-state (ss) $Q$ function of the cavity field, $Q_{\rm ss}(x+iy)=\pi^{-1}\langle x+iy|\rho_{\rm cav, ss}|x+iy \rangle$~\cite{CarmichaelBook1}---we call $\rho_{\rm cav}$ the reduced system density matrix upon tracing out the two-state atom---is commonly used to indicate the coexistence of two {\it quasi}coherent states in certain regions of the drive-amplitude-detuning plane: one close to the vacuum, termed {\it dim} (D), and another of high photon number, termed {\it bright} (B). The domain of state coexistence~\cite{Wellens2001, Ginossar2010, Peano2010, Gabor2023, Sett2024}, which may also be observed in a multiphoton resonance output~\cite{Mavrogordatos2022, Mavrogordatos2024}, shapes the path via which photon blockade breaks down by means of a first-order dissipative quantum phase transition~\cite{Vojta2003, Carmichael2015, Fink2017, Vukics2019, Brookes2021, Kurko24} or via an order-to-disorder transition~\cite{Karmstrand24}. The first operational example we meet in Sec.~\ref{subsec:cohlocJC} illustrates low-photon bistable switching, conditioned on a particular sequence of recorded photon ``clicks'' in the domain of state coexistence, for an equated dwell time in the B and D states, amounting to tens of cavity lifetimes. 

At the same time, bistability is also evinced by the solution to the MBEs, which are obtained by taking expectation values in Heisenberg equations of motion and factorizing the expectations of operator products. The MBEs in the limit of ``zero system size'' with finite detuning, corresponding to the ME~\eqref{eq:ME}, read~\cite{Carmichael2015}
\begin{subequations}\label{eq:MBEs}
\begin{align}
\frac{d\alpha}{dt}&=-(\kappa-i\Delta\omega)a-ig\beta-i\varepsilon,\\
\frac{d\beta}{dt}&=i\Delta\omega \beta + ig\alpha\zeta,\\
\frac{d\zeta}{dt}&=2ig(\alpha^{*}\beta-a\beta^{*}),
\end{align}
\end{subequations}
where $\alpha=\langle a \rangle$, $\beta=\langle \sigma_{-} \rangle$ and $\zeta=\langle \sigma_z \rangle=2\langle \sigma_{+}\sigma_{-}\rangle-1$. They conserve the length of the Bloch vector, $4|\beta|^2 + \zeta^2=1$. Their solution nonlinearly depends on $\Delta \omega/\kappa$, $g/\kappa$ as well as on $\varepsilon/\kappa$ through the system excitation. 

Setting to zero the LHS of the MBEs~\eqref{eq:MBEs}, we find that the semiclassical (neoclassical) photon amplitude in the steady state satisfies the nonlinear equation
\begin{equation}\label{eq:neoclassical}
|\tilde{\alpha}_{\rm ss}|^2=\left(\frac{2\varepsilon}{g}\right)^2 \left[1 + \left( \frac{|\Delta \omega|}{\kappa}-\frac{1}{\sqrt{\Delta\omega^2 \kappa^2/g^4 + |\tilde{\alpha}_{\rm ss}|^2}}\right)^2 \right]^{-1},
\end{equation}
where $|\tilde{\alpha}_{\rm ss}|^2\equiv(|\alpha_{\rm ss}|^2/n_{\rm scale})$ is the scaled intracavity excitation. Solutions to the state equation~\eqref{eq:neoclassical}, in addition to the B and D states of semiclassical amplitude bistability, also indicate the presence of a third state, which is {\it unstable} (U) to fluctuations~\cite{Lugiato1984, Savage1988, CarmichaelBook2, Carmichael2015} and leaves no trace of a peak in the ensemble average over many realizations, visualized by a steady-state $Q$ function or a photon-number distribution. Nevertheless, as we will find in Sec.~\ref{sec:cohlocinv}, {\it conditioned} phase-space distributions of the cavity field evince one, or perhaps a few, short-lived unstable-state peaks. 

Equation~\eqref{eq:neoclassical}, in spite of neglecting quantum correlations, serves another important purpose: its form explicitly suggests that the cavity excitation is scaled by the parameter $n_{\rm scale} \equiv [g/(2\kappa)]^2$, defining a strong-coupling ``thermodynamic limit''~\cite{Carmichael2015, Dombi2015}. When quantum dynamics enter the description, they produce a continuous disagreement with the semiclassical predictions as $n_{\rm scale}\to \infty$, with the persistence of photon blockade being a notable example in the regime $|\Delta \omega|\sim g$~\citep{Carmichael2015, Mavrogordatos2019}. Moreover, closer to resonance where a qualitative agreement between the ME and semisclassical predictions is attained, quantum fluctuations shrink the regime of $Q$-function bimodality with respect to that of semiclassical bistability, as illustrated in Fig. 2(a) of~\citep{Carmichael2015}.

Our principal aim in this paper is to uncover the role of the unstable state in the bistable switching, unraveling ME~\eqref{eq:ME} into individual quantum trajectories~\cite{Carmichael1993QTI, Carmichael1993QTII, Carmichael1997} and gaining access to the conditional state of the cavity field. A clear motivation for appealing to conditional {\it quasi}probability distributions to describe the ergodic fluctuations of a monitored quantum source dates back to the inception of quantum trajectories on the basis of photoelectric detection theory~\cite{Mandel1958,KelleyKleiner1964,Davidson1968,Saleh1978,SrinivasDavies1981,Mandel1981,SrinivasDavies1982,Carmichael1993QTI}. An illustrative example is provided by Figure 10.8 of~\cite{Carmichael1993QTIV}, depicting the generation of Schr\"{o}dinger cat states during a switch initiated by an atomic collapse in phase bistability~\cite{Alsing1991, Kilin91, Mabuchi2009}. 

\section{Coherent localization and its inversion}
\label{sec:cohlocinv}

\subsection{Null-measurement record for an empty cavity}
\label{subsec:nullempty}

Let us now come to our first example of a record with no photon ``clicks'', or a so-called null-measurement record, unraveling the ME~\eqref{eq:ME} with no drive ($\varepsilon=\omega_d=0$) and with the two-state system decoupled from the cavity mode ($g=0$). In a frame rotating with the resonant mode frequency, this leaves us with
\begin{equation}\label{eq:MEdm}
\frac{d\rho}{dt}\equiv \mathcal{L}_{\kappa}\rho=\kappa(2a\rho a^{\dagger}-a^{\dagger}a\rho-\rho a^{\dagger}a).
\end{equation}
In other words, we consider an empty cavity mode prepared in a coherent state $|\alpha\rangle$---a continuous deterministic evolution with initial condition $|\overline{\psi}_{\rm REC}(0)\rangle=|\alpha\rangle$, decaying to the vacuum. When registering a photoelectron counting record for a damped coherent state with complex amplitude $\alpha(t)=\alpha e^{-\kappa t}$, the un-normalized conditioned state is (see Appendix~\ref{subsubsec:dampedA} for more details)
\begin{equation}
|\overline{\psi}_{\rm REC}(t)\rangle=\lambda(\tfrac{1}{2}|\alpha|^2;t)|\alpha(t)\rangle,
\end{equation}
and the probability density for a null measurement result in the interval $[0,t)$ is
\begin{equation}
\langle \overline{\psi}_{\rm REC}(t)|\overline{\psi}_{\rm REC}(t) \rangle=\lambda(|\alpha|^2;t),
\end{equation}
where~\cite{Carmichael2013Ch4}
\begin{equation}
\lambda(x;t)\equiv \exp[-x(1-e^{-2\kappa t})].
\end{equation}
 
\begin{figure*}
\centering
\includegraphics[width=\textwidth]{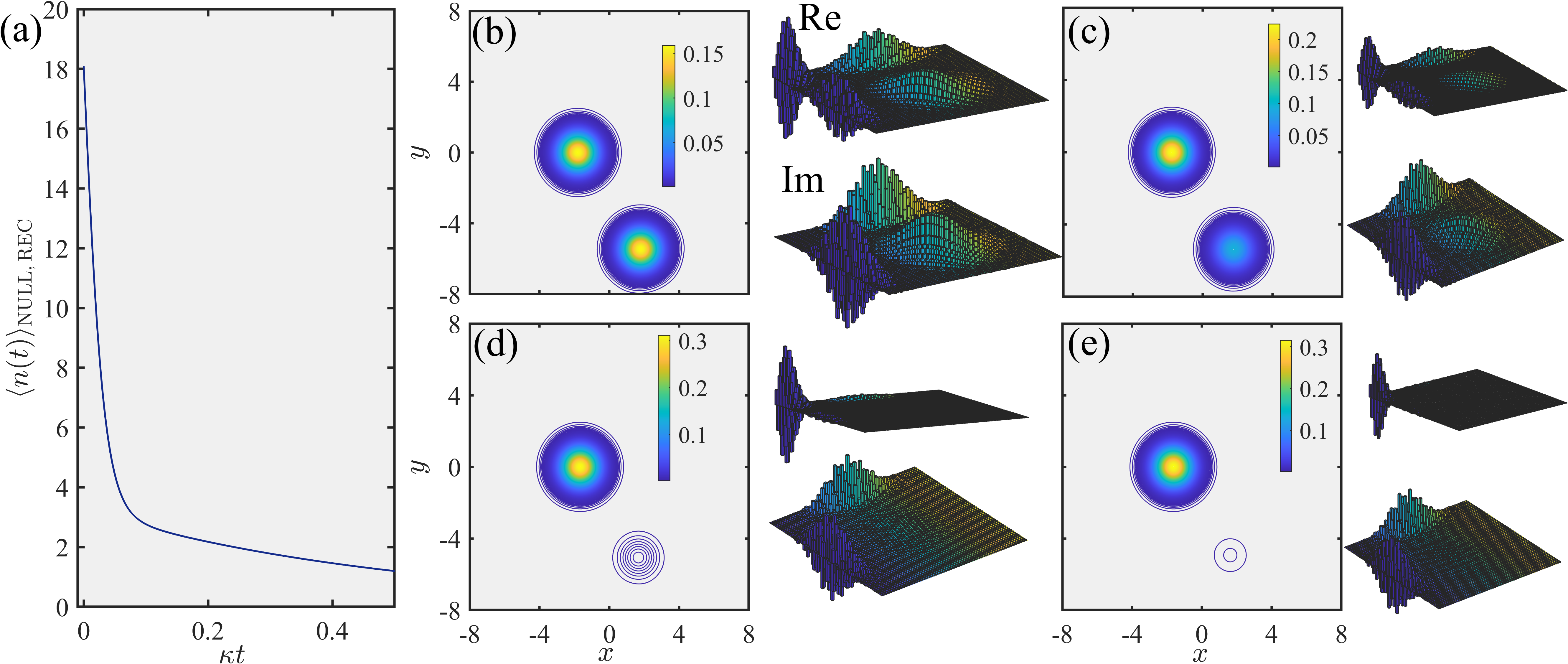}
\caption{{\it Tracking a null-measurement record.} {\bf (a)} Conditioned photon number  $\langle n(t)\rangle_{\rm REC,\, NULL}$ against the dimensionless time $\kappa t$ along a sample trajectory unraveling ME~\eqref{eq:MEdm}, i.e., for a damped mode in an empty cavity, initialized at the state of Eq.~\eqref{eq:inputStatemain} with $\alpha_1=1.8-5.45i$ and $\alpha_2=-1.8$. Frames {\bf (b)--(e)} depict contour plots of the conditioned $Q$ function $Q_{\rm REC}(x+iy;t_k)$ corresponding to $\langle n(t_k)\rangle_{\rm REC,\, NULL}=18.09, 11.64, 3.24$ and $2.77$, for $k=1{\rm (b)}, 2{\rm (c)}, 3{\rm (d)}$ and $4{\rm (e)}$, respectively, with reference to frame (a). To the right of each frame are shown unlabelled barplots of the real (Re) and imaginary (Im) parts of the corresponding conditioned density matrix $(\rho_{\rm cav;\, REC}(t_k))_{mn}$. The Fock-state basis is truncated at $L_{\rm max}=60$.}
\label{fig:FIG1}
\end{figure*}

A more intricate situation arises when we consider the decay of a coherent-state superposition~\cite{Buzek1992, SGComment1994, Carmichael1994, Carmichael1999, Glancy08, Deleglise2008} in a damped resonator with initial state~\cite{Carmichael2013Ch4}
\begin{equation}\label{eq:inputStatemain}
|\overline{\psi}_{\rm REC}(0)\rangle=\frac{1}{\sqrt{2}}\frac{|\alpha_1\rangle + |\alpha_2\rangle}{1+{\rm Re}\langle \alpha_1|\alpha_2\rangle},
\end{equation}
where $\alpha_1$ and $\alpha_2$ are two complex numbers which are to be identified with the peak location of the bright and unstable states, respectively, in individual trajectories unraveling ME~\eqref{eq:ME} when the dipole coupling and drive are turned on. The initial photon number in the superposition state~\eqref{eq:inputStatemain} is
\begin{equation}\label{eq:nN0}
\langle n(t=0)\rangle_{\rm REC}=\frac{|\alpha_1|^2 + |\alpha_2|^2 + 2{\rm Re}(\alpha_{1}^{*} \alpha_{2} \langle\alpha_1|\alpha_2\rangle)}{2[1+{\rm Re}(\langle \alpha_1|\alpha_2\rangle)]}.
\end{equation}
Under the no ``click'' evolution (null-measurement record) for fixed coherent-state amplitudes, the evolution is again deterministic and continuous, and the conditional photon number at time $t$ is given by:
\begin{equation}\label{eq:ntexact}
\langle n(t)\rangle_{\rm REC,\, NULL}=\frac{\langle \overline{\psi}_{\rm REC,\, NULL} (t)|a^{\dagger}a|\overline{\psi}_{\rm REC,\, NULL} (t)\rangle}{\langle \overline{\psi}_{\rm REC,\, NULL} (t)|\overline{\psi}_{\rm REC,\, NULL} (t)\rangle},
\end{equation}
a deterministic nonlinear function of the exponentially decaying coherent-state amplitudes $\alpha_1(t)=\alpha_1 e^{-\kappa t}$, $\alpha_2(t)=\alpha_2 e^{-\kappa t}$. As shown in the Appendix~\ref{subsubsec:dampedcs}, when $|\alpha_1-\alpha_2|^2\gg 1$, the conditioned photon number is approximated to very good accuracy by:
\begin{equation}\label{eq:napp}
\langle n(t) \rangle_{\rm REC,\, NULL}=\frac{|\alpha_1(t)|^2\lambda(|\alpha_1|^2;t) + |\alpha_2(t)|^2\lambda(|\alpha_2|^2;t)}{\lambda(|\alpha_1|^2;t)+\lambda(|\alpha_2|^2;t)}.
\end{equation}
The conditional photon-number average depicted in Fig.~\ref{fig:FIG1}(a) is indistinguishable from the approximation of Eq.~\eqref{eq:napp} on the scale of the figure. The barplots of the conditional distributions of the intracavity field reveal persisting off-diagonal portions of the state representation, even when the high-amplitude state peak has virtually vanished: in that case depicted in Fig.~\ref{fig:FIG1}(e), we are still dealing with a superposition of coherent states with complex amplitudes.

In the sections that follow, we will be modelling a quantum jump in amplitude bistability with a coherent-state superposition as an input state to an empty cavity. The initial density matrix in the Fock-state basis at $t=0$ is 
\begin{equation}
\begin{aligned}
&(\rho_{\rm cav}(t=0)_{\rm REC,\,NULL})_{mn}\equiv\langle m |\rho_{\rm cav}(t=0)_{\rm REC, \,NULL}| n \rangle\\
&=\frac{\left(2[1+{\rm Re}(\langle \alpha_1|\alpha_2\rangle)] \right)^{-1}}{\sqrt{m!n!}}\sum_{i=1}^{2}\sum_{j=1}^{2}e^{-\frac{1}{2}(|\alpha_i|^2+|\alpha_j|^2)}\alpha_i^{*m}\alpha_j^n.
\end{aligned}
\end{equation}
The last two terms distinguish a pure coherent-state superposition from the mixed state that solves ME~\eqref{eq:ME} in the steady state. Hence, in the Fock-state basis, the matrix elements of the initial density matrix contain the terms $\propto (\sqrt{n!m!})^{-1}e^{-\frac{1}{2}|\alpha_1|^2}e^{-\frac{1}{2}|\alpha_2|^2}\left(\alpha_1^{*m}\alpha_2^n+\alpha_2^{*m}\alpha_1^n\right)$, corresponding to two squeezed off-diagonal distributions peaking at about $(m,n)=(|\alpha_1|^{2}, |\alpha_2|^{2})$ and its diagonally symmetric location, with an unequal width along the rows and columns, $\sim 2|\alpha_1|, 2|\alpha_2|$. They indicate a coherent superposition, as opposed to the statistical mixture solving the master equation in the steady state. The latter mixed steady states correspond to the {\it quasi}probability distributions $Q_{\rm ss}(x+iy)$ depicted {\it e.g.}, in the inset of Figs.~\ref{fig:FIG2}(a), \ref{fig:FIG3}(f). 

The conditional cavity density matrix during a null-measurement record with initial state~\eqref{eq:inputStatemain} reads
\begin{equation}\label{eq:rhoCt}
\begin{aligned}
(\rho_{\rm cav}(t)_{\rm REC,\,NULL})_{mn}&\equiv\langle m |\rho_{\rm cav}(t)_{\rm REC, \,NULL}| n \rangle\\
=\frac{C^{-1}(t)}{\sqrt{m!n!}}&\sum_{i=1}^{2}\sum_{j=1}^{2}\sqrt{\lambda(|\alpha_i|^2;t)\lambda(|\alpha_j|^2;t)}\\
&\times e^{-\frac{1}{2}(|\alpha_i(t)|^2+|\alpha_j(t)|^2)}\alpha_i^{*m}(t)\alpha_j^n(t),
\end{aligned}
\end{equation}  
where 
\begin{equation*}
\begin{aligned}
C(t)\equiv& \lambda(|\alpha_1|^2;t)+\lambda(|\alpha_2|^2;t)\\
&+2\lambda(|\alpha_1|^2/2;t)\lambda(|\alpha_2|^2/2;t)\,{\rm Re}(\langle\alpha_1(t)|\alpha_2(t)\rangle).
\end{aligned}
\end{equation*}
The last two terms in square brakets in Eq.~\eqref{eq:rhoCt}, responsible for the off-diagonal distributions in Figs.~\ref{fig:FIG1}(b--e), are particularly sensitive to the complex amplitudes of two states between which localization takes place. For instance, changing the amplitudes $\alpha_1, \alpha_2$ in Fig.~\ref{fig:FIG1} to real numbers equal to their magnitudes $|\alpha_1|, |\alpha_2|$, would significantly reduce the maximum of the off-diagonal distributions in ${\rm Re}[(\rho_{\rm cav}(t_e)_{\rm REC,\,NULL})_{mn}]$ with respect to those seen in ${\rm Im}[(\rho_{\rm cav}(t_e)_{\rm REC,\,NULL})_{mn}]$ for the same instance of time $t_e$. When the full JC nonlinearity is brought into play, the amplitudes $\alpha_1$ and $\alpha_2$ are themselves stochastic quantities, read off from different conditional cavity distributions in the course of the quantum jumps recorded along individual realizations. We infer that the manifestation of coherence in the process of localization is sensitive to the position of the bright ($\alpha_1$) and unstable ($\alpha_2$) states in  phase space, as well as to their fluctuations across different B$\to$D jumps.  Different peak locations in phase space, conditioned on a particular photoelectron counting record, give rise to different off-diagonal distributions. Cavity-state tomograms~\cite{Deleglise2008,Hofheinz2008,Eichler2012} may then operationally track the de-excitation path along the JC energy ladder---see, {\it e.g.}, Fig. 3(b) of~\cite{Carmichael2015}---or the displacement of the unstable-state peak in phase space, as we will find in Sec.~\ref{subsec:cohlocJC} [{\it e.g.}, in Figs.~\ref{fig:FIG3}(c--e)].

\subsection{Coherent localization in the Jaynes--Cummings downward switching events of bistability}
\label{subsec:cohlocJC}

At this stage, we turn on the dipole coupling strength $g$ and the external drive $\varepsilon$, to deal with ME~\eqref{eq:ME} for the open coherently driven JC oscillator. Based on the null-measurement evolution for an empty cavity developed in Sec.~\ref{subsec:nullempty}, the prescription to model a particular jump by the empty-cavity coherent localization along a sample record $\langle n(t)\rangle^{\rm JCD}_{\rm REC}$ (see Appendix~\ref{sec:MC}) includes the following steps: 

{\bf (i)} Read off the two coherent state amplitudes of the bright (B) and unstable (U) states, denoted by $\alpha_1$ and $\alpha_2$, respectively, from the conditional {\it quasi}probability distribution.

{\bf (ii)} Calculate the initial photon number $\langle n (t=0)\rangle_{\rm REC}$ from Eq.~\eqref{eq:nN0} and determine the abscissa $t_{\rm mid}$ corresponding to the ordinate $\langle n(t=t_{\rm mid})\rangle^{\rm JCD}_{\rm REC}=\langle n(t=0)\rangle_{\rm REC}$, which moves the time origin from $0$ to $t_{\rm mid}$, different for every jump.

{\bf (iii)} Graphically locate the time $t^{\prime}\equiv (t_{\rm mid}+\Delta t_{\rm end})$ at which $\langle n(t=t^{\prime})\rangle_{\rm REC,\, NULL}=|\alpha_2|^2$.

{\bf (iv)} Invert {\it by hand} the plot of $\langle n(t)\rangle_{\rm REC,\, NULL}$ about the point with co-ordinates $(t_{\rm mid},\langle n(t=0) \rangle_{\rm REC})$ for a time interval equal to $\Delta t_{\rm end}$. 

\begin{figure*}
\centering
\includegraphics[width=\textwidth]{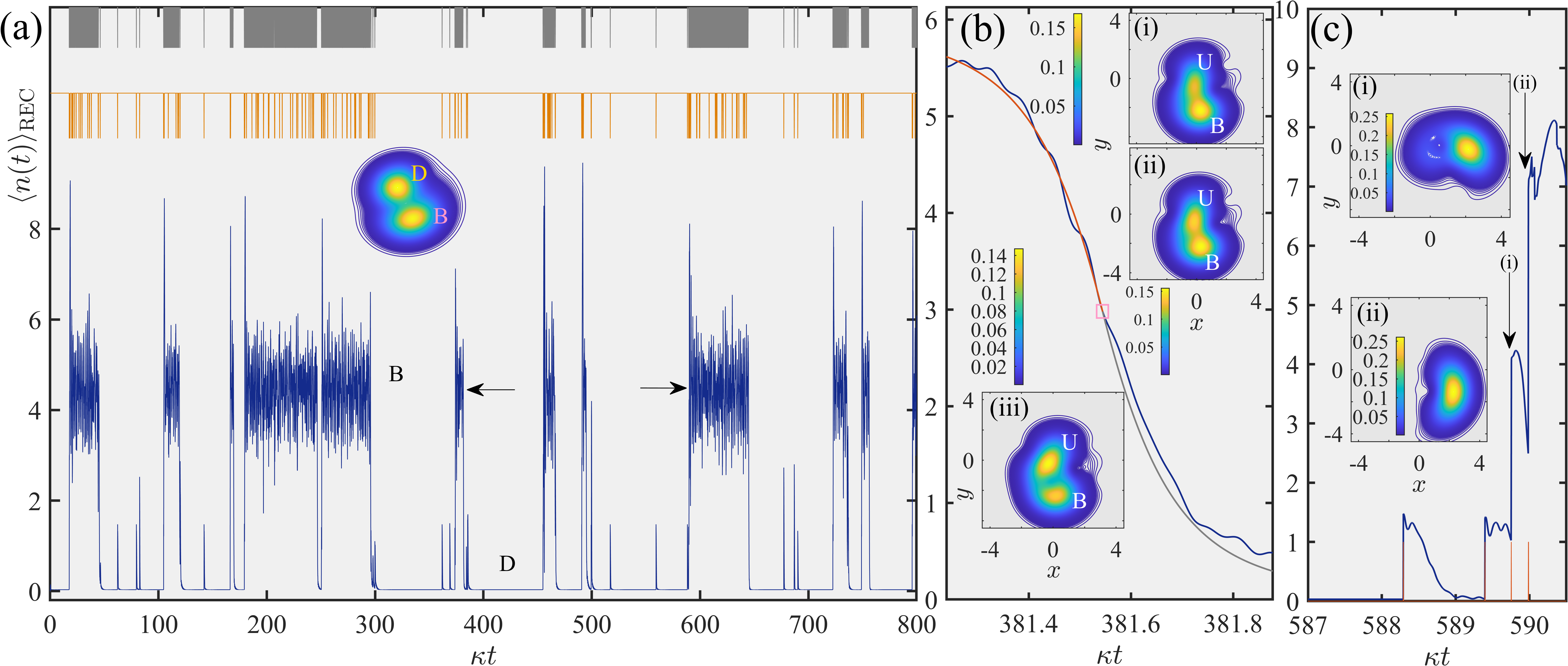}
\caption{{\it Low-amplitude quantum bistable switching.} {\bf (a)} Sample realization of the conditioned photon number $\langle n_{\rm REC}(t)\rangle$ for $800$ (average) cavity lifetimes, showing bistable switching between the two metastable states, B and D. The upper (lower) row of vertical strokes indicates photon emissions (upward jumps). The inset depicts a schematic contour plot of $Q_{\rm ss}(x+iy)$(see Appendix~\ref{sec:MC} and Ref.~\cite{Tan1999}). {\bf (b)} Focus on the continuous segment of the B$\to$D jump plotted against the dimensionless time and modelled by a coherent localization which takes place between a bright state (B) with amplitude $\alpha_1=0.3-2.38i$ and an unstable state (U) with amplitude $\alpha_2=-0.175-0.53i$. The average conditioned photon number $\langle n(t)\rangle^{\rm JCD}_{\rm REC}$ is plotted in blue, the part of $\langle n(t)\rangle_{\rm REC,\,NULL}$ [Eq.~\eqref{eq:ntexact}] from $t_{\rm mid}$ to $t^{\prime}\equiv(t_{\rm mid}+\Delta t_{\rm end})$ is plotted in grey, while the time-inverted part of $n_{\rm REC,\,NULL}(t)$ from $t_{\rm mid}$ to $t^{\prime \prime}\equiv(t_{\rm mid}-\Delta t_{\rm end})$ is plotted in orange. The pink square marks the inversion point. The inset plots the entire realization of $\langle n(t)\rangle_{\rm REC}$ for $800$ cavity lifetimes, where the jump B$\to $D is indicated by the left-pointing arrow in (a). The three insets (i--iii) depict contour plots of $Q_{\rm REC}(x+iy;t_k)$ corresponding to $\langle n(t_k)\rangle^{\rm JCD}_{\rm REC}=3.02, 2.92$ and $2.70$, for $k=1{\rm (i)}, 2{\rm (ii)}, 3{\rm (iii)}$, respectively. {\bf (c)} Focus on the jump D$\to $B indicated by the right-pointing arrow in (a). Orange strokes mark upward jumps (the pair of emissions at $\kappa t \approx 590$ is not resolved on the scale of the figure). The two insets (i, ii) depict contour plots of the conditional distribution $Q_{\rm REC}(x+iy;t_k)$ corresponding to $\langle n(t_k)\rangle^{\rm JCD}_{\rm REC}=4.07$ and $7.08$, for $k=1 {\rm (i)}, 2{\rm (ii)}$, respectively. B, D, U, denote the bright, dim and unstable states, respectively. The operating parameters read: $g/\kappa=25$, $\varepsilon/\kappa=5.3$ and $\Delta\omega/\kappa=-8$.}
\label{fig:FIG2}
\end{figure*}

\begin{figure*}
\centering
\includegraphics[width=\textwidth]{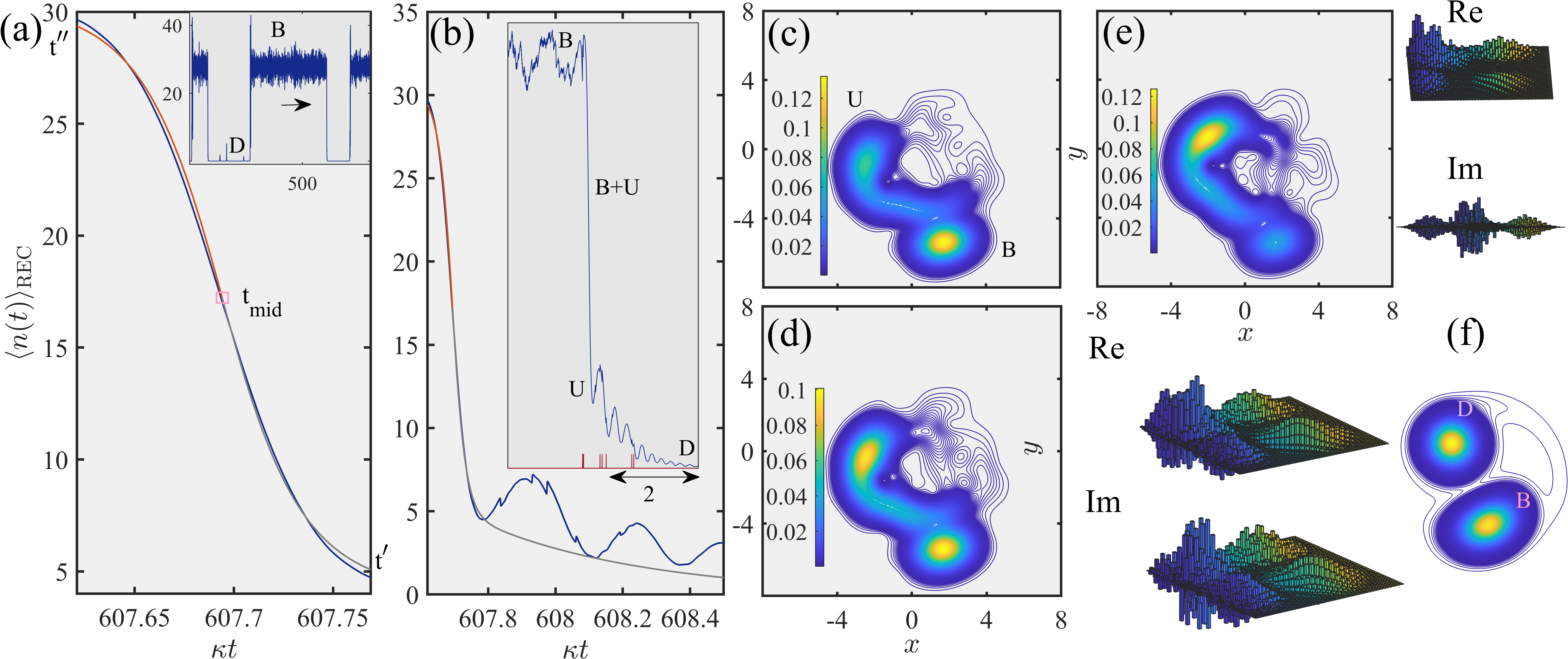}
\caption{{\it State localization in high-amplitude bistability.} {\bf (a)} Focus on the continuous segment of the B$\to$D jump, plotted against the dimensionless time and modelled by a coherent localization which takes place between a bright state with amplitude $\alpha_1=1.7-5.15i$ and an unstable state with amplitude $\alpha_2=-2.25-0.2i$. The average conditioned photon number $\langle n(t)\rangle^{\rm JCD}_{\rm REC}$ is plotted in blue, the part of $\langle n(t)\rangle_{\rm REC,\,NULL}$ [Eq.~\eqref{eq:ntexact}] from $t_{\rm mid}$ to $t^{\prime}$ is plotted in grey, while the time-inverted part of $n_{\rm REC,\,NULL}(t)$ from $t_{\rm mid}$ to $t^{\prime \prime}$ is plotted in orange. The pink square frames the inversion point. The inset depicts the corresponding realization during $800$ cavity lifetimes in which the B$\to$ D jump in question is indicated by the arrow. {\bf (b)} Broader perspective of the localization B$\to$U in question, continued beyond $t^{\prime}=(t_{\rm mid}+\Delta t_{\rm end})$ to show the decay to D. The inset zooms out even further and also indicates the up jumps (red strokes) as well as the scale of an average photon lifetime. Dominant states are marked along the jump. {\bf (c--e)} Contour plots of $Q_{\rm REC}(x+iy;t_k)$, corresponding to $\langle n(t_k)\rangle^{\rm JCD}_{\rm REC}=19.39, 16.51$ and $11.19$, for $k=1, 2, 3$ in frames (c, d, e), respectively, with reference to frame (a). Schematic barplots of the real (Re) and imaginary (Im) parts of the corresponding conditioned cavity density matrix $(\rho_{\rm cav;\, REC}(t_k))_{mn}$ are drawn for the times $t_2, t_3$, to the right of frames (d, e), respectively. {\bf (f)} Schematic contour of the steady-state distribution $Q_{\rm ss}(x+iy)$. B, D, U, denote the bright, dim and unstable states, respectively. The operating parameters read: $g/\kappa=60$, $\varepsilon/\kappa=13.5$ and $\Delta\omega/\kappa=-8$.}
\label{fig:FIG3}
\end{figure*}

Thus we reach the time $t^{\prime\prime}=t_{\rm mid}-\Delta t_{\rm end}$. This is the path taken to produce the plots of Figs.~\ref{fig:FIG2}(b) and \ref{fig:FIG3}(a). 


Figure~\ref{fig:FIG2}(a) depicts a typical realization of low-amplitude bistable switching with a steady-state average photon number $\langle a^{\dagger}a \rangle_{\rm ss}\approx 2.46$, similar to the kind presented in~\cite{Dombi2015}, or monitored in~\cite{Kerckhoff11} via heterodyne detection~\cite{Schack1997, BreuerPetruccione}---see also Sec~\ref{subsec:MQ}. Steady-state emission is bunched and the zero-delay autocorrelation~\cite{CarmichaelBook2} is $\langle (a^{\dagger})^2a^2 \rangle_{\rm ss}/(\langle a^{\dagger}a \rangle_{\rm ss})^2\approx 1.75$. For the twelve switching events to the bright state shown, upward jumps~\cite{Chough2001} (those leading to an increase in $\langle n(t)\rangle^{\rm JCD}_{\rm REC}$; see also Appendix~\ref{sec:MC}) amount to about $10\%$ of the total cavity emissions. The localization shown in Fig.~\ref{fig:FIG2}(b) is completed in a time window $2\kappa \Delta t_{\rm end}\approx 0.638$, and the analytical expression~\eqref{eq:ntexact}---together with its inversion about the photon number~\eqref{eq:nN0}---are in good agreement with the conditional photon counting record $\langle n(t)\rangle^{\rm JCD}_{\rm REC}$ produced by the Monte Carlo algorithm~\cite{Dalibard1992, Dum1992, Molmer1996, Stenholm1997, Plenio1998, Vogel2006, BreuerPetruccione, Carmichael2013Ch4}---see also Appendix~\ref{sec:MC}. The conditional cavity distributions drawn in the insets of Fig.~\ref{fig:FIG2}(b) provide a visualization of the coherent localization in terms of the bright and unstable states with time-varying peak heights in the $Q$ function. In contrast to B$\to$D jumps, D$\to$B jumps {\it must be} conditioned on upward collapses. As we observe in Fig.~\ref{fig:FIG2}(c), the corresponding photon emissions are highly bunched, occurring at a rate comparable to the light-matter coupling strength $g$. Squeezed coherent states~\cite{Mandel1982, Walls1983, Reid1985, Lugiato1989, Agasti2024} are conditioned on the upward collapses, following an excitation path built along the rung of the JC dressed states~\cite{Carmichael2015}. 

\begin{figure*}
\centering
\includegraphics[width=\textwidth]{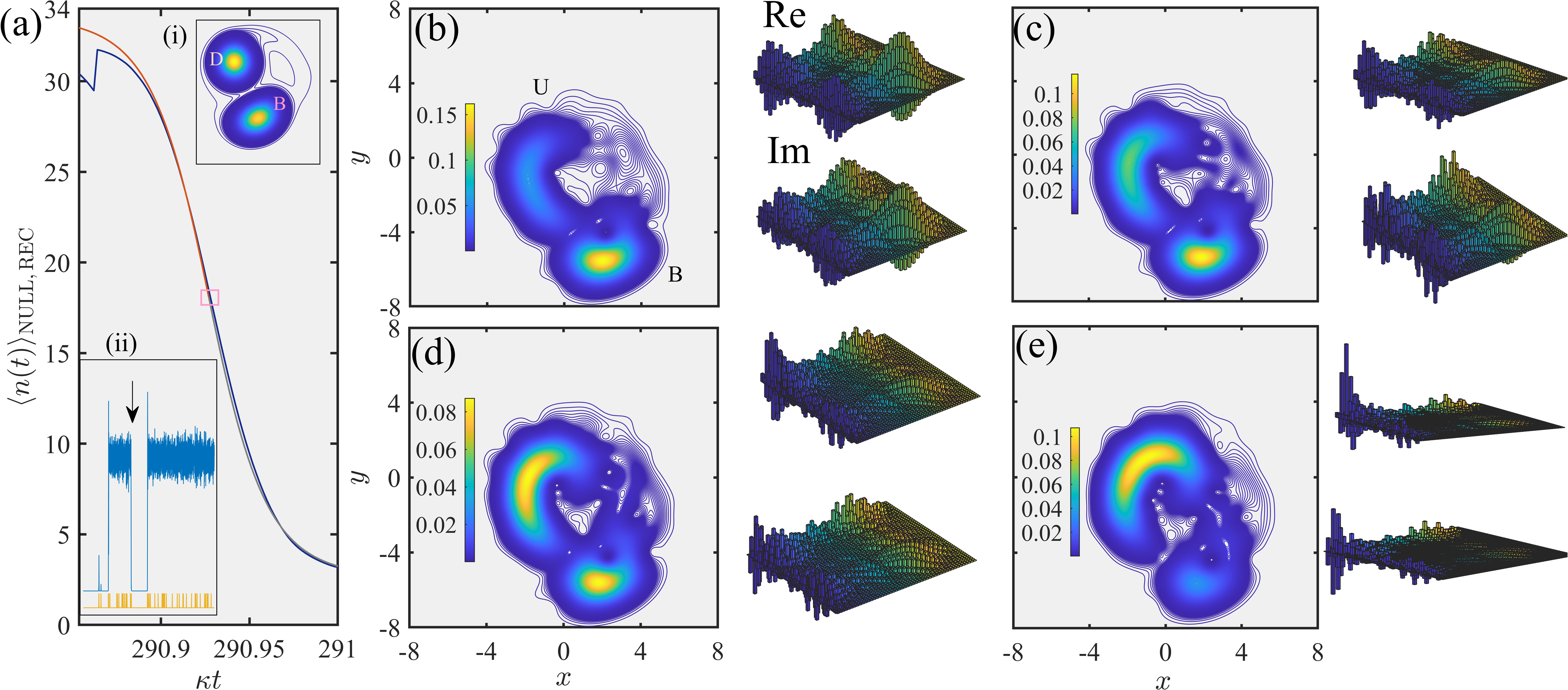}
\caption{{\it An alternative localization between fluctuating states.} {\bf (a)} Focus on the {\it quasi}continuous segment of the B$\to$D jump (interrupted only by one cavity emission) in $\langle n(t)\rangle^{\rm JCD}_{\rm REC}$, plotted against the dimensionless time and modelled by a coherent localization which takes place between a bright state with amplitude $\alpha_1=1.8-5.45i$ and an unstable state with amplitude $\alpha_2=-1.8$. The average conditioned photon number $\langle n(t)\rangle^{\rm JCD}_{\rm REC}$ is plotted in blue, the part of $\langle n(t) \rangle_{\rm REC,\,NULL}$ [Eq.~\eqref{eq:ntexact}] from $t_{\rm mid}$ to $(t_{\rm mid}+t_{\rm end})$ is plotted in grey, while the time-inverted part from $t_{\rm mid}$ to $(t_{\rm mid}-t_{\rm end})$ is plotted in orange. The pink square frames the inversion point. Inset (i) plots a schematic contour of the steady-state Q function $Q_{\rm ss}(x, y)$, and inset (ii) depicts the corresponding realization during $800$ cavity lifetimes in which the B$\to$ D jump in question is indicated by the arrow. The strokes underneath indicate the photon emissions corresponding to upward jumps. {\bf (b--e)} Contour plots of the conditioned distributions $Q_{\rm REC}(x+iy;t_k)$ corresponding to $\langle n(t_k)\rangle^{\rm JCD}_{\rm REC}=23.06, 18.04, 14.60$ and $7.80$, for $k=1, 2, 3, 4$ in frames (b, c, d, e), respectively, with reference to frame (a). Schematic barplots of the real (Re) and imaginary (Im) parts of the corresponding conditioned cavity density matrix $(\rho_{\rm cav;\,REC}(t_k))_{mn}$ are drawn to the right of each frame. The operating parameters read: $g/\kappa=60$, $\varepsilon/\kappa=13.5$ and $\Delta\omega/\kappa=-8$. The Fock-state basis is truncated at $L_{\rm max}=60$.}
\label{fig:FIG4}
\end{figure*}

\begin{figure*}
\centering
\includegraphics[width=\textwidth]{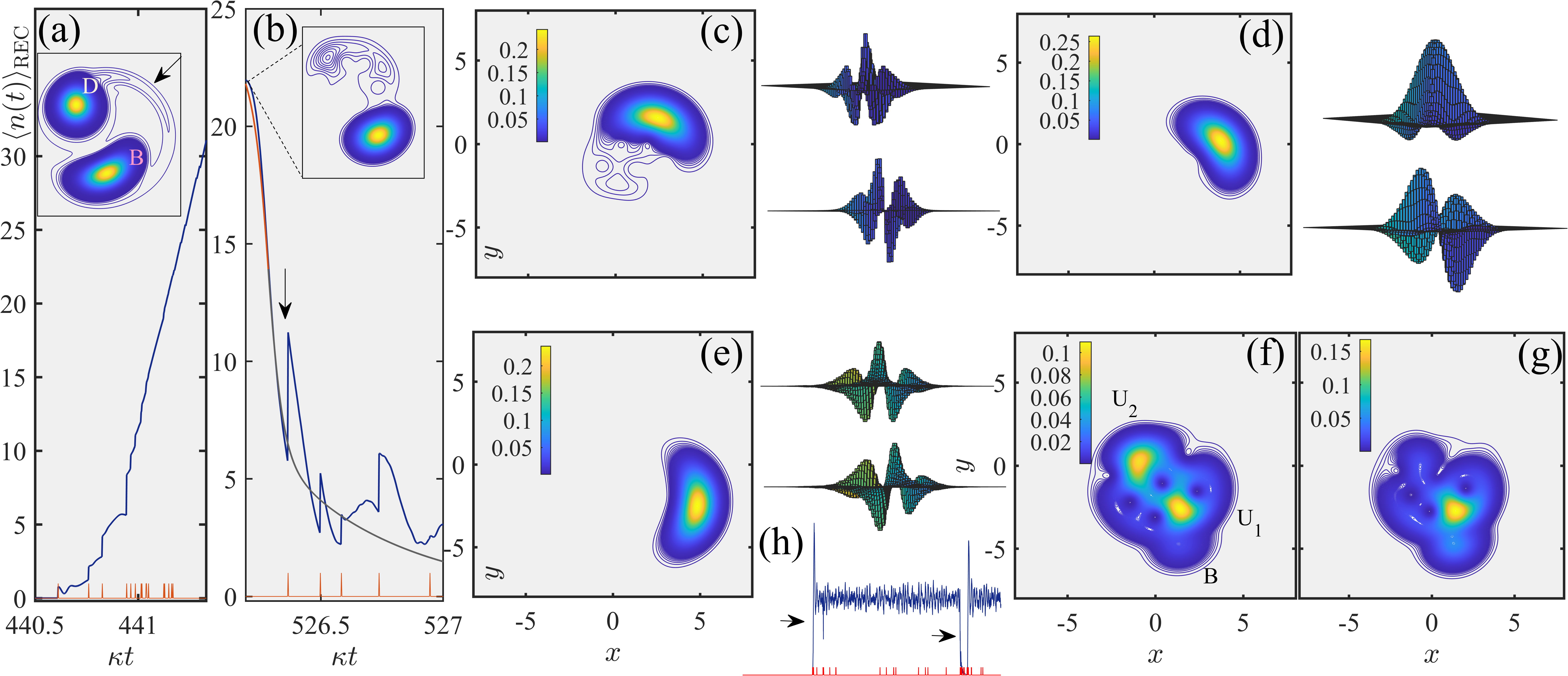}
\caption{{\it Upward jumps in the bistable switching.} {\bf (a)} Conditioned photon number $\langle n(t)\rangle^{\rm JCD}_{\rm REC}$ focused on a fraction of a D$\to$B jump for $0.83$ cavity lifetimes, and indicated by the left arrow in (h). Orange strokes mark upward jumps. The framed inset plots a schematic contour of the steady-state Wigner function $W_{\rm ss}(x, y)$, and the arrow marks the excitation path. {\bf (b)} Focus on the B$\to$D jump for $0.808$ cavity lifetimes, indicated by the right arrow in (h). The plot of Eq.~\eqref{eq:ntexact} is overlaid in gray, modelling a coherent localization between a bright state with amplitude $\alpha_1=2-4.2i$ and an unstable state with amplitude $\alpha_2=-1.95+1.55i$. The inverted part with respect to $\langle n(\kappa t_{\rm mid}=526.285)\rangle_{\rm REC,\, NULL}$ is plotted in orange, reaching $|\alpha_1|^2=21.64$. The framed inset depicts the schematic contour of the conditional $Q$-function corresponding to the maximum of $\langle n(t)\rangle^{\rm JCD}_{\rm REC}$. {\bf (c--e)} Conditioned distributions $Q_{\rm REC}(x+iy;t_k)$ corresponding to $\langle n(t_k)\rangle^{\rm JCD}_{\rm REC}=8.37, 15.20$ and $29.31$, for $k=1, 2, 3$ in frames (c, d, e), respectively, with reference to the D$\to$B jump shown in frame (a). Schematic barplots of the real (Re) and imaginary (Im) parts of the corresponding conditioned cavity density matrix $(\rho_{\rm cav;\, REC}(t_k))_{mn}$ are drawn to the right of each {\it quasi}probability distribution. {\bf (f--g)} Contour plots of the conditioned distributions $Q_{\rm REC}(x+iy;t_k)$ corresponding to $\langle n(t_k)\rangle^{\rm JCD}_{\rm REC}=5.77$ and $11.08$, for $k=1, 2$ in frames (f, g), respectively, with reference to the B$\to$D jump shown in frame (b). These conditioned photon averages are attained immediately before [$\langle n(t_k)\rangle^{\rm JCD}_{\rm REC}=5.77$] and after [$\langle n(t_k)\rangle^{\rm JCD}_{\rm REC}=11.08$] the upward jump indicated by the arrow in (b). {\bf (h)} The inset plots the entire trajectory of $800$ cavity lifetimes, where the D$\to$B jump of (a) is indicated by the left arrow and the the B$\to$D jump in (b) is indicated by the right arrow. Red strokes mark upward jumps. The operating parameters read: $g/\kappa=50$, $\varepsilon/\kappa=11.6$ and $\Delta\omega/\kappa=-8$. The Fock-state basis is truncated at $L_{\rm max}=50$.}
\label{fig:FIG5}
\end{figure*}

Let us now move to high-amplitude bistability for $\langle a^{\dagger}a \rangle_{\rm ss}\approx 14.65$, and examine the mechanism of quantum bistable switching to the dim state. For trajectories in these operating conditions, upward jumps make less or about $1\%$ of the total photon emissions; however, the steady-state autocorrelation is $\langle (a^{\dagger})^2a^2 \rangle_{\rm ss}/(\langle a^{\dagger}a \rangle_{\rm ss})^2\approx 1.89$, evincing a higher photon bunching than the low-amplitude bistability of Fig.~\ref{fig:FIG2}(a). The localization depicted in Fig.~\ref{fig:FIG3}(a) is completed in a considerably shorter time, $2\kappa \Delta t_{\rm end}\approx 0.149$, and is in an even better agreement with the analytically obtained  null-measurement evolution and its inversion. The off-diagonal distributions in the conditional cavity density matrix in a Fock-state basis, complementing the $Q$ functions in Figs.~\ref{fig:FIG3}(c--e), testify to a persisting coherent-state superposition in all stages of the probability transfer. For the localization illustrated in Fig~\ref{fig:FIG3}(a), Eq.~\eqref{eq:jumptime} gives $\kappa \Delta t_{\rm end,\,min} \approx 0.051$. Substituting into the same expression the steady-state amplitudes of semiclassical (sc) bistability $|\alpha_1|_{\rm ss,\, sc} \approx 5.30$ and $|\alpha_2|_{\rm ss,\,sc} \approx 2.08$, obtained from Eq.~\eqref{eq:neoclassical}, gives $\kappa \Delta t_{\rm end,\,min,\ sc} \approx 0.053$.

Along a single realization, a B$\to $D jump is anticipated by a short interval of monotonically increasing $\langle n(t)\rangle^{\rm JCD}_{\rm REC}$. While the increase in the conditional photon number---signalling the appearance of the unstable state and lasting a time comparable to the localization that is to follow---is in principle continuous and monotonic, it may as well be interrupted by cavity emissions. The emissions immediately prior to the localization, however, occur at a much slower rate than those conditioning the bright state. In that interval, the bright state increases in magnitude while a connecting path is being established to the unstable state in the phase space, to which coherent localization is to take place. Once the unstable state has formed, we are in position to identify the amplitudes $\alpha_1$ (B) and $\alpha_2$ (U) in step (i) of our prescription.  

From Eq.~\eqref{eq:napp}, we can infer the time duration required for the coherent localization which, when completed, yields $\langle n (\Delta t_{\rm end})\rangle_{\rm REC,\, NULL}=|\alpha_2|^2$, {\it i.e.}, the excitation of the unstable state . After inversion about the mid-point is performed, we expect to reach the bright state with $\langle n(t=t^{\prime}) \rangle_{\rm REC,\, NULL}=|\alpha_1|^2$. The total time duration for the localization is then $2\Delta t_{\rm end}$, where $t_{\rm end}$ is determined by the graphical intersection of $|\alpha_1(t)|^2\lambda(|\alpha_1|^2;t) + |\alpha_2(t)|^2\lambda(|\alpha_2|^2;t)$ and $|\alpha_2|^2 [\lambda(|\alpha_1|^2;t)+\lambda(|\alpha_2|^2;t)]$. The order of magnitude of $\kappa \Delta t_{\rm end}$ depends not only on the ratio of the well separated coherent state amplitudes, but is also sensitive to the modulus of the unstable-state amplitude $|\alpha_2|$. For the coherent localization illustrated in Fig.~\ref{fig:FIG3}, we find an intersection at $\kappa \Delta t_{\rm end} \approx 0.0743$. For $\kappa \Delta t_{\rm end} \ll 1$, a condition upheld in the strong-coupling limit for ``zero system size'', the following expression yields a lower bound for the localization duration $2 \Delta t_{\rm end}$, identifying (see Appendix~\ref{subsubsec:dampedcs}):
\begin{equation}\label{eq:jumptime}
\boxed{
\Delta t_{\rm end,\,min}\approx -\frac{\ln\left(\displaystyle\frac{|\alpha_2|}{|\alpha_1|^2-|\alpha_2|(|\alpha_2|-1)}\right)}{2\kappa (|\alpha_1|^2-|\alpha_2|^2)}.}
\end{equation} 
The substitution $|\alpha_{1,2}|\to |\alpha_{1,2}|_{\rm ss,\,sc}$ in the above formula provides an estimate for the average localization time over an ensemble of trajectories, taking into account the Poisson fluctuations of the photon number about the mean of the unstable state. In this sense, the accuracy of the under-estimated lower bound increases for $n_{\rm scale} \to \infty$ and higher drive amplitude in the bistability regime, where relative photon fluctuations about the mean for the bright state vanish (see also the progression between frames in Fig. 4 of~\cite{Carmichael2015}). Equation~\eqref{eq:jumptime} predicts a longer localization for a bright state closer in magnitude to the unstable, while the indeterminacy arising for $|\alpha_1|\to |\alpha_2|$ points to the special case of phase bistability and the formation of even/odd cat states. 

Is is instructive to consider the special case of bistable switching in the strong-coupling limit at the right end of the semiclassical hysteresis ``S'' curve~\cite{Savage1988, CarmichaelBook2}, where the bright state has a large amplitude $|\alpha_1|=A \gg 1$, while the unstable state is formed with a photon content of the order of unity ($|\alpha_2|\sim 1$). This property can be verified by the solutions of the state equation~\eqref{eq:neoclassical}. In this case, the jump from the bright to the dim state effectively coincides with the coherent localization from the bright to the unstable state, since the dim and unstable states are both well separated from the bright in phase space, and practically indistinguishable in a background of sub-photon fluctuations. Equation~\eqref{eq:jumptime} predicts then the following asymptotic behaviour for the downward switching time:
\begin{equation}\label{eq:asympt}
\kappa \Delta t_{\rm loc,\,min} \equiv \kappa \Delta t_{\rm switch\, B\to D}\sim \frac{\ln A^2}{A^2},
\end{equation}
describing its approach to zero for $A^2 \to \infty$. As we move towards the left edge end of the hysteresis curve, the unstable state gains an appreciable excitation $|\alpha_2|$ and is well separated from the dim state in phase space; the dominant correction to the jump time is given by the subtraction of $\ln|\alpha_2|$ from the numerator of Eq.~\eqref{eq:asympt}. Overall, the application of Eq.~\eqref{eq:jumptime} with $|\alpha_{1,2}|\to |\alpha_{1,2}|_{\rm ss,\,sc}$ shows that the localization time increases towards the left edge of the ``S'' curve, when the drive amplitude is lowered together with the intracavity excitation.

We can also place the new timescale in context with reference to the decoherence rate of cat states, by considering a damped cavity mode prepared to the initial state $|\psi_{\rm REC,\, NULL}(0)\rangle$ [Eq.~\eqref{eq:inputStatemain}] with $\alpha_2=-\alpha_1=A$---a large-amplitude even cat state~\cite{HarocheBook, Carmichael2013Ch4}. The change over time from a Wigner function for the density matrix [obeying ME~\eqref{eq:ME} with $\varepsilon=g=0$] which initially shows interferences, to a diagonal form of a statistical mixture of coherent states proceeds extremely quickly if the initial coherent state amplitude $A$ is large on the scale of a single quantum. We observe that coherent-state localization, between a highly excited bright state with $A\gg 1$ and an unstable state with mean photon number about unity, is predicted to last longer by a factor $\sim \ln A$ compared to the decoherence time $\sim (2\kappa A^2)^{-1}$ waited on average for the loss of the very first photon from the damped cavity mode initialized in the cat state. Hence, a quantum jump between two macroscopic states of light {\it along a single trajectory} unraveling the ME of quantum bistability involves an intermediate timescale in between the energy damping time $(2\kappa)^{-1}$ and the decoherence time $(2\kappa A^2)^{-1}$ of a damped coherent-state superposition with the bright-state excitation, quantities pertaining to an {\it ensemble average over individual realizations}. 

In fact, the two aforementioned scenarios are markedly different with regard to the ensemble-average evolution: critical slowng down~\cite{Dykman2012, Brookes2021} to the steady state in quantum bistability is to be contrasted with the rapid decoherence time of a damped cavity mode initialized in a cat state. Moreover, at the level of an ensemble average of an observable like the cavity photon number, the intermediate timescale in question is not tractable nor is the unstable state distribution. It is however interesting to remark that the localization time in individual jumps does not depend on the weight of the bright and dim components in the steady-state statistical mixture $\rho_{\rm ss}$ solving ME~\eqref{eq:ME}. This weight primarily affects critical slowing down as well as the lifetime of the two metastable states (B and D) alternating in the individual realizations, both of which may exceed the energy damping time by several orders of magnitude~\cite{Carmichael2015, Sett2024}.

Let us stay a bit more with the indeterminacy noted in Eq.~\eqref{eq:jumptime} when $|\alpha_1|\to |\alpha_2|$. Apart from Schr\"{o}dinger cat states, notable cases of interest are the neoclassical states of phase bistability ($\alpha_1=\alpha_2^{*}$). From an operational perspective, the competition between the two states $|\overline{\psi}^{(\alpha_1)}_{\rm REC}(t) \rangle$ and $|\overline{\psi}^{(\alpha_2)}_{\rm REC}(t) \rangle$, evolving continuously between jumps registered at the ordered time sequence $t_1,t_2,...t_n$ (with $n$ counts up to $t$) as
\begin{equation}
\begin{aligned}
&|\overline{\psi}^{(\beta)}_{\rm REC}(t) \rangle=(\sqrt{2\kappa}\,\beta\,e^{-\kappa t_n})\ldots (\sqrt{2\kappa}\,\beta\,e^{-\kappa t_1})\\
&\times \exp\left[-\tfrac{1}{2}|\beta|^2 (1-e^{-2\kappa t})\right]\,|\beta e^{-\kappa t}\rangle, \quad \beta=\alpha_1, \alpha_2,
\end{aligned}
\end{equation}
to dominate their superposition, no longer occurs (see Appendix~\ref{subsubsec:dampedcs}). Can we establish a connection between these two speccial cases from the same operational viewpoint? In~\cite{Carmichael1993QTIV}, we find that Schr\"{o}dinger cats are generated during a phase switch initiated by an atomic collapse. Following a spontaneous emission---a single registered microscopic event---the conditioned $Q$ function splits into four peaks. Two of them, corresponding to emissions into the sidebands of the Mollow spectrum, sweep through each other as they go through a phase switch, and ultimately reassemble with the other two, corresponding to emissions into the central peak of the Mollow spectrum. The coherent evolution depicted in Fig. 10.8 of~\cite{Carmichael1993QTIV} lasts about $1.8\kappa^{-1}$, with $\alpha_{1,2}\approx 8\pm 5i$, {\it i.e.}, an order of magnitude longer than the coherent localization for the bistable output depicted in Fig.~\ref{fig:FIG3}. 

There is yet another dynamical aspect which makes the complex-conjugate states of phase bistability special. In~\citep{Alsing1991}, we find that these neoclassical states, obtained when taking $\gamma/(2\kappa)\to 0$ to the MBEs and then producing a solution, organize the asymptotic semiclassical dynamics. In contrast, the solutions to the MBEs with $\gamma$ finite, exist only in the formal sense and are approached by a rate vanishingly small as $\gamma/(2\kappa)\to 0$. The limit of ``zero system size'' is thus structurally unstable, and the solutions to the MBEs with $\gamma/(2\kappa)\to 0$  are {\it a priori} fragile to fluctuations; however, they become ``attractors'' in the full quantum-mechanical treatment. Reinstating spontaneous emission and breaking the conservation law of the Bloch-vector magnitude~\cite{Alsing1991, CarmichaelBook2, Carmichael2015} renders the complex-conjugate states unstable to fluctuations in the quantum dynamics. In absorptive bistability, they only appear as a pair of unstable states during switching event to the low-photon attractor, as we will later see in Fig.~\ref{fig:FIG12} of Sec~\ref{sec:SE}. The pair is short-lived and fuses to a single state which subsequently decays to the dim state. Traces of the semiclassical dynamics beyond steady-state bistability can be found in individual quantum trajectories. 

From the mean-field equations we may extract two dominant frequencies which, for the regime of semiclassical bistability considered in this work, feature as an envelope and a carrier; the frequency of the former is an order of magnitude smaller than $g$. A regression of that sort is pictured in Fig.~\ref{fig:FIG3}(b), following the localization `fluctuation' at about $\kappa t=607.8$ when the unstable state has dominated the cavity distribution. The beat `envelope' is still tied to the null-measurement evolution of the preceding localization, while the waveform is interrupted by jumps (photon emissions) until the dim state is reached. Therefore, the concept of a semiclassical ringing, alongside its merge with a fast oscillation $\sim g$, follows us from the unraveling of a two-photon resonance peak, as we will find in Sec.~\ref{sec:mpJC}, up to the quantum jumps of amplitude bistability. In the latter, the fast oscillation effectively sets the measure for the time intervals between the upward collapses conditioning the D$\to$B jump. 

Let us track a B$\to$D jump occurring for the same operating conditions as in Fig.~\ref{fig:FIG3}, but along a different trajectory. As we might expect, the amplitudes $\alpha_1$ and $\alpha_2$ identified from the conditional $Q$-functions in Fig.~\ref{fig:FIG4} are now different since the states involved are subject to fluctuations; they are actually the same with those selected for the initial state in the null-measurement record unraveling ME~\eqref{eq:MEdm} in Fig.~\ref{fig:FIG1}. Nevertheless, coherent localization occurs in exactly the same fashion and gives rise to persisting off-diagonal distributions across all stages of the probability transfer. Moreover, using the graphical intersection method to solve $\langle n(t) \rangle_{\rm REC, NULL}=|\alpha_2|^2$ with the magnitudes $|\alpha_1|, |\alpha_2|$ provided in the caption of Fig.~\ref{fig:FIG4}, yields $\kappa \Delta t_{\rm end}\approx 0.073$, almost identical to the localization time $\kappa \Delta t_{\rm end}\approx 0.074$ calculated for the jump depicted in Fig.~\ref{fig:FIG3}. This instance suggests that while $\alpha_1$ and $\alpha_2$ fluctuate from one B$\to$D jump to another as well as across jumps in different trajectories, their magnitudes eventually lead to virtually identical localization times. We recall here that the derivation of Eq.~\eqref{eq:jumptime} only takes into account fluctuations in the unstable state and therefore systematically underestimates the localization time.  

Figure~\ref{fig:FIG5} aims to compare a D$\to$B to a B$\to$D switch with respect to the role of upward jumps. The two events we focus on form part of a trajectory of $800$ cavity lifetimes, where the upward jumps only make less than $1\%$ of the total cavity emissions. The main difference can be directly spotted: while those jumps are a necessary condition to a D$\to$B switching event, for the B$\to$D switch they work against coherence. The conditional cavity distributions plotted in frames (c--g) support this view, while visualizing the fundamentally different role of quantum fluctuations underlying the two processes. 

Similar to Fig.~\ref{fig:FIG2}(c), in Fig.~\ref{fig:FIG5}(a) we see that a D$\to$B jump is conditioned on a bundle of upward collapses at a rate competing with the light-matter coupling strength $g$ [compare also with the transition from Fig. 4(a) to Fig. 4(b) in~\cite{Chough2001}]. Squeezed coherent states spiral towards higher excitation after such jumps, climbing up the rung of dressed JC states~\cite{Carmichael2015} while following the excitation path marked by the arrow in the inset of Fig.~\ref{fig:FIG5}(a). Traces of coexistent states are observed in the conditional $Q$ function depicted in Fig.~\ref{fig:FIG5}(c); however, these are not accompanied by the corresponding off-diagonal distributions of a coherent-state superposition. The remaining frames visualizing the conditional increase in intracavity excitation evince the density matrix of a squeezed coherent state with its peak tracing the excitation path in phase space. The path is reflected in the varying proportions of positive and negative-valued regions in the Fock-state resolution of the pure cavity states. 
\begin{figure*}
\centering
\includegraphics[width=\textwidth]{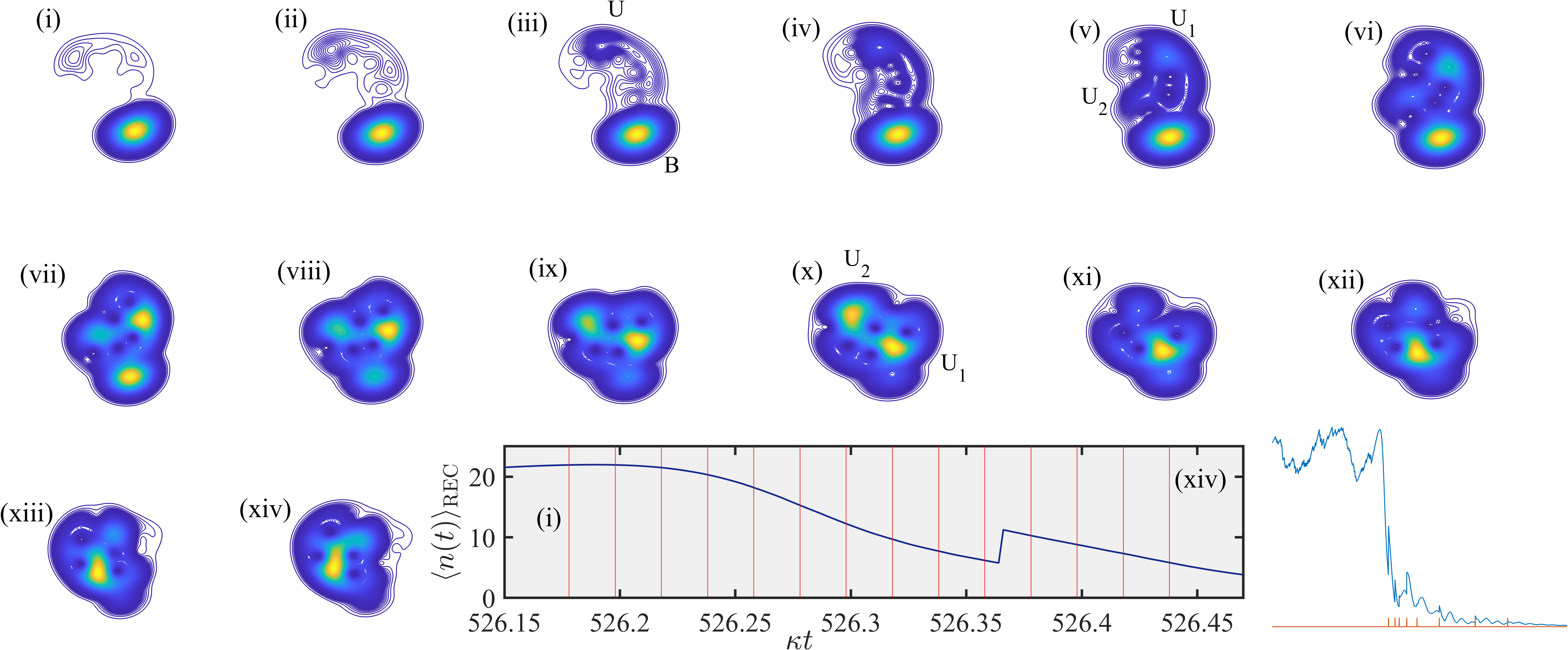}
\caption{{\it Splitting and spiralling of the unstable state.} Schematic contour plots of the conditioned distributions $Q_{\rm REC}(x+iy;t_k)$ corresponding to the $k=1 {\rm (i)}, 2{\rm (ii)}\ldots 14{\rm (xiv)}$ conditional photon averages marked by the red lines intersecting the B$\to$D jump [identical to that depicted in Fig~\ref{fig:FIG5}(b)] shown in the bottom frame: $\langle n(t_k)\rangle^{\rm JCD}_{\rm REC}=21.93, 21.91, 21.49, 20.32, 18.20, 15.28, 12.24$, $9.69, 7.72, 6.19, 10.28, 8.79, 7.33$ and $5.83$. A broader perspective of the B$\to$D jump is given at the bottom right for $6$ cavity lifetimes. Orange strokes mark upward jumps. The operating parameters read: $g/\kappa=50$, $\varepsilon/\kappa=11.6$ and $\Delta\omega/\kappa=-8$. The Fock-state basis is truncated at $L_{\rm max}=50$.}
\label{fig:FIG6}
\end{figure*}
\begin{figure*}
\centering
\includegraphics[width=\textwidth]{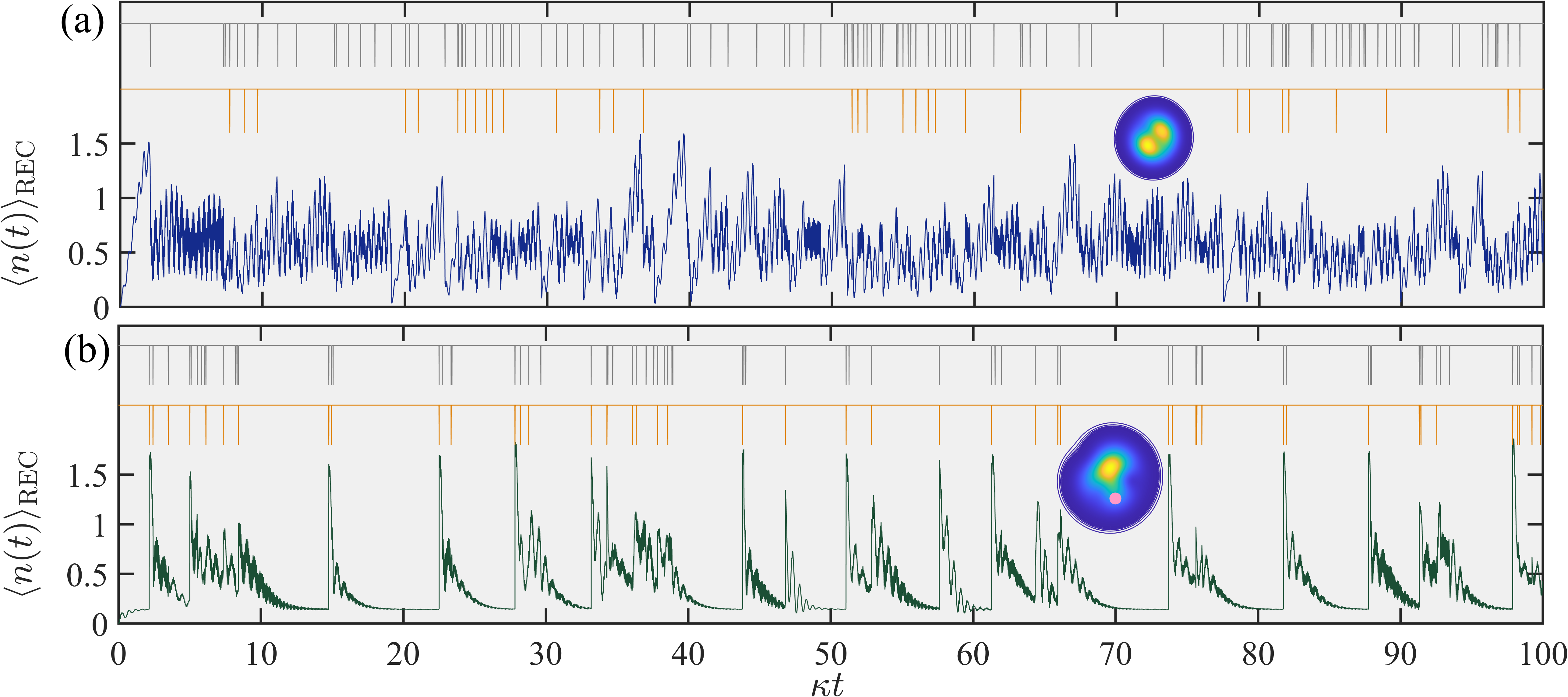}
\caption{{\it Multiphoton resonances and jump statistics.} Conditioned photon number  $\langle n(t)\rangle^{\rm JCD}_{\rm REC}$ along a sample trajectory unraveling ME~\eqref{eq:ME} for $100$ cavity lifetimes. The upper row of gray strokes indicates all types of cavity emissions, while the lower row in orange marks upward jumps. The operating parameters are $g/\kappa=50$, $\varepsilon/\kappa=5.3$ and $\Delta \omega/\kappa=36.10$ in {\bf (a)} (two-photon resonance peak); $\Delta \omega/\kappa=29.35$ in {\bf (b)} (three-photon resonance peak). The steady state average photon number $\langle a^{\dagger} a \rangle_{\rm ss}$ and zero-delay intensity correlation function $g^{(2)}(0)=\langle a^{\dagger\,2} a^2 \rangle_{\rm ss}/\langle a^{\dagger} a \rangle_{\rm ss}^2$, extracted from the numerical solution of the ME~\eqref{eq:ME}, are: $0.56$ and $0.85$, respectively, for (a); $0.39$ and $2.31$, respectively, for (b). Record (a) registers $122$ photon emissions (``clicks''), out of which $32$ correspond to upward jumps, while record (b) registers $74$ photon emissions out of which $45$ correspond to upward jumps. The schematic contour plots in both frames depicts the steady-state Wigner distribution $W_{\rm ss}(x+iy)$ of the cavity mode. The Fock-state basis is truncated at $L_{\rm max}=20$.}
\label{fig:FIG7}
\end{figure*} 

The picture drastically changes in Figs.~\ref{fig:FIG5}(f--g) devoted to the B$\to$D switch. In frame (f) we observe a complex of unstable states, $U_1$ and $U_2$, resolved into a single peak by an upward jump in frame (g) transferring some probability to the bright state. The peak that remains ($U_1$) is the one corresponding to the state with which coherent localization with the bright state took place beforehand. Otherwise, the upward jump disturbs the null-measurement evolution but is not enough to prevent the localization, on the basis of the interplay we describe in Appendix~\ref{subsubsec:dampedcs}.

A more complete `conditional jump tomography' is depicted in Fig.~\ref{fig:FIG6} for a fraction of the evolution we saw in Fig.~\ref{fig:FIG5}(b). Fourteen contour plots show the transition from the initial appearance of the unstable state coexisting with the bright state, to the rotation of the unstable state in phase space before it splits into a complex of states. Localization involves one of these peaks, as we can observe in frames (vii) and (viii). Frame (xi) depicts the resolution of the complex to the state mediating the coherent localization, following one upward jump -- the same instance we saw in Figs.~\ref{fig:FIG5}(f--g).   

\section{Multiphoton Jaynes--Cummings resonances: a precursor of amplitude bistability}
\label{sec:mpJC}

As a further application of the Monte Carlo algorithm, let us focus on two sample realizations generated for conditions of two- and three-photon blockade--imagine a line of constant $\varepsilon/\kappa=5.3$ cutting through the lower panel of Fig. 1 in~\citep{Carmichael2015}. For the unraveling at the two-photon resonance peak shown in Fig.~\ref{fig:FIG7}(a), we can notice a bimodal phase-space profile in the Wigner function of the steady-state intracavity field~\cite{CarmichaelBook1}. We can also observe the so-called separation of timescales reported in~\cite{Mavrogordatos2024}, where a semiclassical oscillation of frequency $\sim\varepsilon^2/(g\kappa)$ is set apart from a quantum beat at $\approx 2g$ by means of single photon emission events. Steady-state photon emission is anti-bunched and, expectedly, less than half of the total collapses (about $26\%$) correspond to upward jumps. 

A qualitative difference now sets in for the photoelectron counting record obtained at the three-photon peak and depicted in Fig.~\ref{fig:FIG7}(b), despite the fact that both resonances attain a maximum intracavity excitation of less than one photon in steady state. There is a visible accumulation of upward jumps to the remnant of a high-photon state marked by the pink dot in the Wigner function. There is also a clear relaxation to the dominant attractor---a state of $0.15$ photons corresponding to the peak in the Wigner function. Upward jumps make the $61\%$ of the total, which is again expected by the rise of the autocorrelation at zero delay to a value above $2$. Finally, a high-frequency modulation of the conditional emission probability is still visible, but here it reflects an interference of quantum beats~\cite{Mavrogordatos2022}. 

What basis of comparison can be formed between the conditioned photon record at the three-photon resonance peak and the trajectory of low-photon bistable switching we saw in Fig.~\ref{fig:FIG2}? First, there is a tendency of upward collapses to precipitate jumps to a higher-photon state. A similar number of such jumps occur between the two records. However, in the case of amplitude bistability, several highly bunched upward jumps are required to establish the bright state, in contrast to a single jump producing a high-amplitude fluctuation along the three-photon resonance record. 

Single jumps are also registered as occasional fluctuations off the dim state in a trajectory of amplitude bistability, but they rapidly decay within only half of an average cavity lifetime. The time it takes for single-emission fluctuations to regress seems to be a key difference between the two operating conditions. At a multiphoton resonance, the regression of fluctuations is dominated by a slower, semiclassical timescale, linked to the saturation of an effective two-level transition~\cite{Shamailov2010, Mavrogordatos2022, Mavrogordatos2024}. Traces of the semiclassical ringing timescale, dictated by the $n$-photon effective Rabi frequency, disappear as photon blockade breaks down. On the other hand, it is the fast regression timescale in amplitude bistability that requires multiple upward jumps to establish the bright (B) metastable state, instead of an occasional fluctuation. Figure~\ref{fig:FIG5}(a) clearly exemplified this case, where tens of upward jumps are recorded in a duration shorter than the average cavity lifetime, before the switch to the bright state is completed. We will comment more on this instance in Sec~\ref{sec:photonstat}.

\begin{figure*}
\centering
\includegraphics[width=\textwidth]{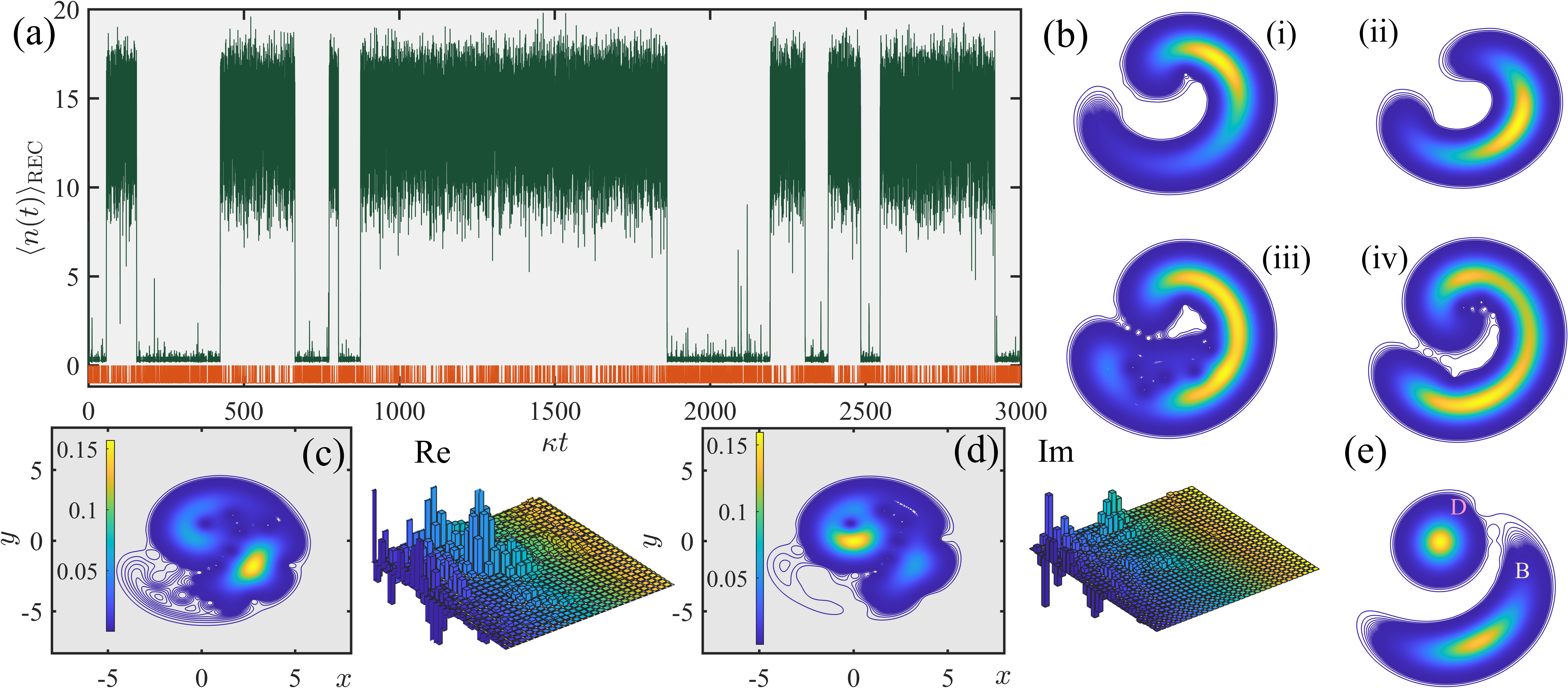}
\caption{{\it Instability to fluctuations under the Kerr nonlinearity.} {\bf (a)} Conditional average photon number $\langle n(t) \rangle^{\rm Kerr}_{\rm REC}$ along a sample trajectory unraveling ME~\eqref{eq:MEKerr} for $3\times10^3$ cavity lifetimes. Strokes underneath indicate upward jumps. {\bf (b)} Schematic contour plots of the conditional cavity distribution $Q_{\rm REC}(x+iy;t_k)$, corresponding to $\langle n(t_k)\rangle^{\rm Kerr}_{\rm REC}=13.74, 15.58, 15.55$ and $16.59$, for $\kappa \,t_k=875.1300, 875.2050, 664.1925$ and $804.2250$ where $k=1$ (i), $k=2$(ii), $k=3$(iii) and $k=4$(iv), respectively, with reference to frame (a). {\bf (c--d)} Contour plot of the conditional cavity distribution $Q_{\rm REC}(x+iy;t_k)$ corresponding to $\langle n(t_k)\rangle^{\rm Kerr}_{\rm REC}=8.12$ and $4.34$ for $\kappa \,t_k=2915.5655$ and $2915.6850$ where $k=1$(c) and $2$(d), respectively, with reference to frame (a). The schematic barplots to the right of frames (c) and (d) depict the real and imaginary parts, respectively, of the conditional cavity density matrix elements $(\rho_{\rm cav;\, REC}(t_k))_{mn}$, $k=1, 2$, corresponding to the $Q$ functions on the left. {\bf (e)} Steady-state Wigner distribution of the cavity matrix, $W_{\rm ss;\,Kerr}(x,y)$, numerically calculated via exact diagonalization of the Liouvillian in ME~\eqref{eq:MEKerr}, in agreement with the analytical expression~\eqref{eq:Wssexact}. The operating parameters read: $\kappa/\chi=2$, $\varepsilon/\chi=16.5$ and $\Delta\omega/\chi=20$. The steady-state photon number and zero-delay autocorrelation are $\langle a^{\dagger}a \rangle_{\rm ss}\approx 8.11$ and $\langle a^{\dagger 2}a^2 \rangle_{\rm ss}/\langle a^{\dagger}a\rangle_{\rm ss}^2\approx 1.42$. D, B mark the dim, bright states respectively. The Fock-state basis is truncated at $L_{\rm max}=31$ and the time step used in the Monte Carlo algorithm is $\kappa\Delta t=0.0075=0.15(\chi/\Delta\omega)$.}
\label{fig:FIG8}
\end{figure*}

\section{Quantum jumps in amplitude bistability under the Kerr nonlinearity}
\label{sec:Kerr}

Let us briefly discuss the quantum jumps in the presence of a `milder' anharmonicity than that of the JC spectrum, namely the nonlinearity of the Kerr oscillator which allows for the derivation of analytical ensemble averages in the steady state~\cite{WallsKerr}. In the rotating frame, the ME for the density matrix $\rho$ of the cavity mode is
\begin{equation}\label{eq:MEKerr}
\begin{aligned}
\frac{d\rho}{dt}=&\Delta \omega[a^{\dagger}a,\rho]-i\chi[a^{\dagger 2}a^2,\rho] + \varepsilon[a^{\dagger}-a,\rho]\\
& + \kappa(2a\rho a^{\dagger}-\rho a^{\dagger}a - a^{\dagger}a\rho),
\end{aligned}
\end{equation}
where $\chi$ is a coupling constant for the nonlinear process of self-phase modulation in a medium with third-order susceptibility. The steady-state Wigner function has been derived in~\cite{KheruntsyanKerr} as an explicitly positive distribution,
\begin{equation}\label{eq:Wssexact}
W_{\rm ss;\, Kerr}(x,y)=N e^{-2(x^2+y^2)} \left|\frac{J_{\lambda-1}\left(\sqrt{-8\tilde{\varepsilon} (x-iy)}\right)}{(x-iy)^{(\lambda-1)/2}} \right|^2,
\end{equation}
where $\lambda\equiv (\kappa-i\Delta\omega)/(i\chi)$, $\tilde{\varepsilon}\equiv \varepsilon/(i\chi)$, $J_{\nu}(z)$ is the Bessel function of the first kind, and $N$ is a normalization constant. In certain regions of $\tilde{\varepsilon}$ and $\Delta\omega/\chi$, the oscillator output displays amplitude bistability, while the size of quantum fluctuations is determined by the system-size parameter $n_{\rm scale, \,Kerr}=(\Delta\omega/\chi)^2$ [there is bistability only for $\Delta\omega/\chi>0$, while the parameters $\lambda, \lambda^{*}$ scale the generalized $P$ distribution~\cite{WallsKerr}]. We choose operating conditions corresponding to Fig. 1(a) of~\cite{KheruntsyanKerr}---driving with $|\tilde{\varepsilon}|=16.5$ generates bimodality in the Wigner function~\eqref{eq:Wssexact}. 

Plotting computed Wigner functions tells us nothing about {\it observed} or observable physics, however~\cite{CarmichaelBook2}. A more interesting question is whether scattering records themselves, such as the one pictured in Fig.~\ref{fig:FIG8}, can reveal information about what conditions the bistable switching. There is still a dearth of upward jumps with respect to the total photon emissions, in spite of being much more numerous in the dim state, correlated with occasional fluctuations as shown in Fig.~\ref{fig:FIG8}(a). Bunched sequences of upward collapses condition a D$\to$B jump too (note the absence of an overshoot associated with a D$\to$B jump as in the JC bistable realizations). The conditional cavity distributions underlying the D$\to$B and B$\to$D jumps, however, are notably different to the ones we met for the JC bistability. Instead of localized individual states or a superposition of localized states, we notice a continuous spiral connecting the dim to the bright state in phase space during a jump. A continuous spiral is set in place during a D$\to$B jump, where a squeezed state climbs up the rungs of the Kerr spectrum~\cite{Miranowicz2013} [see Fig.~\ref{fig:FIG8}(b), (i-ii)], as well as for a B$\to$D jump where the probability transfer is highly delocalized in phase space [see Fig.~\ref{fig:FIG8}(b), (iii-iv)]. Such a difference might be anticipated by the form of the steady-state Wigner function pictured in Fig.~\ref{fig:FIG8}(e): the bright state is significantly squeezed and effectively ``connected'' to the dim, while there is no distinction between an excitation and a de-excitation path as in the JC bistability. For the latter, the two paths can be clearly visualized in phase space, {\it e.g.}, in Figs. 3(b) and 6(b) of~\cite{Carmichael2015}. 

Nevertheless, instances of localization may still arise during a B$\to$D jump, involving a state spontaneously formed along the spiral and playing the role of the unstable state, and a state fast decaying to the dim {\it in parallel} with the localization. Figures~\ref{fig:FIG8}(c, d) show two such competing states in a superposition evinced by the off-diagonal distributions in the simulated conditional cavity tomograms. In the absence of a localized unstable state, we might wonder whether Eq.~\eqref{eq:jumptime}, derived on the basis of a null measurement record, is still of some use. It turns out that if $|\alpha_1|$ and $|\alpha_2|$ are substituted with the peak locations of the bright and dim states, respectively, identified from the bimodal $W_{\rm ss}(x,y)$, then Eq.~\eqref{eq:jumptime} provides rather an {\it upper bound} for the B$\to$D jump times, testifying to the fragility of coherent localization as opposed to the JC bistability. We also observe a similar asymmetry in the upward vs. downward jumps (4\% in Fig.~\ref{fig:FIG8}) to the trajectories unraveling the JC ME~\eqref{eq:ME}. However, the records now evince a more evenly spaced photon counting sequence correlated with more frequent fluctuations out of the dim state [compare with Fig.~\ref{fig:FIG2}(a) for the JC bistability].  
\begin{figure*}
\centering
\includegraphics[width=0.97\textwidth]{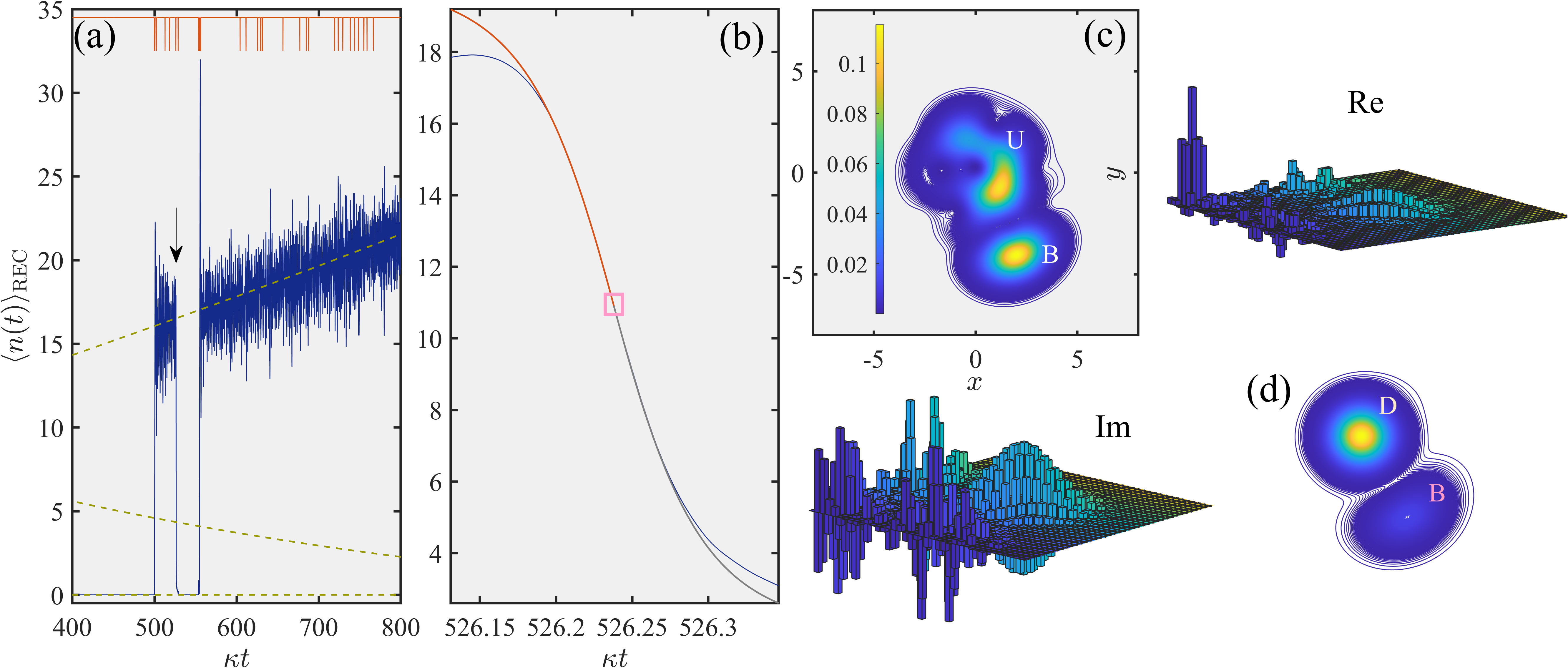}
\caption{{\it Stability against fluctuations, and photon bunching.} {\bf (a)} Conditional average photon number $\langle n(t) \rangle^{\rm JC}_{\rm REC}$ along a sample trajectory unraveling ME~\eqref{eq:ME} with a linearly increasing scaled drive amplitude $\varepsilon/\kappa$ over the range $(0,13]$ along $800$ cavity lifetimes (focus on second half of the interval). The orange strokes on top mark upward jumps. The dashed line depicts the corresponding part of the steady-state neoclassical hysteresis ``S'' curve, after the solution of Eq.~\eqref{eq:neoclassical} for the scaled drive amplitudes in the range $[6.5,13]$---the middle branch is unstable. {\bf (b)} Focus on the continuous segment of the B$\to$D jump, indicated by the arrow in (a) and plotted against the dimensionless time and modelled by a coherent localization which takes place between a bright (B) state with amplitude $\alpha_1=1.9-3.95i$ and an unstable (U) state with amplitude $\alpha_2=1.4-0.8i$. The average conditioned photon number $\langle n(t)\rangle^{\rm JCD}_{\rm REC}$ is plotted in blue, the part of $\langle n(t)\rangle_{\rm REC,\,NULL}$ [Eq.~\eqref{eq:ntexact}] from $t_{\rm mid}$ to $t^{\prime}\equiv(t_{\rm mid}+\Delta t_{\rm end})$ is plotted in grey, while the time-inverted part of $n_{\rm REC,\,NULL}(t)$ from $t_{\rm mid}$ to $t^{\prime \prime}\equiv(t_{\rm mid}-\Delta t_{\rm end})$ is plotted in orange. The pink square marks the inversion point. {\bf (c)} Schematic contour plot of the conditional cavity distribution $Q_{\rm REC}(x+iy;t_c)$, corresponding to $\langle n(t_c)\rangle^{\rm JC}_{\rm REC}\approx 9.41$ for $\kappa \,t_c=526.248$, respectively, with reference to frame (b). Schematic barplots of the real (Re) and imaginary (Im) parts of the corresponding conditioned cavity density matrix $(\rho_{\rm cav;\,REC}(t_k))_{mn}$ are shown on the right and bottom of (c), respectively. {\bf (d)} Schematic contour plot of $Q_{\rm ss}(x+iy)$ obtained from the numerical solution of ME~\eqref{eq:ME} with $\varepsilon=8.5514$--the scaled drive amplitude at $\kappa t_{\rm mid}=526.238$. B and D mark the bright and dim states, respectively. We operate at $g/\kappa=50$, $\Delta\omega/\kappa=-8$, while the Fock-state basis is truncated at $L_{\rm max}=50$.}
\label{fig:FIG9}
\end{figure*}

The Kerr nonlinearity is subject to a weak-coupling limit, one where the system-size parameter tends to zero when a coupling constant modeling the scattering of light by matter (here $\chi$, or $g$ in the JC model) tends to infinity~\cite{Carmichael2015}. Fluctuations are scaled by a $n_{\rm scale, \,Kerr}=(\Delta\omega/\chi)^2=400$, while Figs.~\ref{fig:FIG2} and~\ref{fig:FIG3} have $n_{\rm scale}=156.25$ and $900$, respectively; in contrast to the strong-coupling limit underlying the latter, coherent-state superpositions involving {\it an} unstable state are not robust to fluctuations in the Kerr nonlinearity, and are only spontaneously conditioned in a scattering record. In fact, a mapping to the Kerr oscillator can be made in the dispersive regime of the JC model---where the two-state atomic transition is significantly detuned from the cavity resonance by $\delta \gg g$---setting a basis for a perturbative expansion of the JC square-root nonlinearity~\cite{Bishop2010, Mavrogordatos2016}. The system-size parameter in that case is written as $n_{\rm scale,\, disp.}=[\delta/(2g)]^2 \gg 1$, which also points to a weak-coupling limit.  

\section{Photon correlations, stability against system parameter variation and contextuality of quantum jumps}
\label{sec:photonstat}

A further qualification is in order regarding the role of upward jumps in amplitude bistability, apart from their connection to the photon-number fluctuations in multiphoton resonance operation. As we observed in Sec.~\ref{subsec:cohlocJC}, in the approach to the ``thermodynamic limit'' $n_{\rm scale}=[g/(2\kappa)]^2 \to \infty$, the steady-state autocorrelation constantly increases above unity to approach $2$ in the examples discussed. The ratio of upward jumps to the total, however, tends to zero in a given trajectory. At the same time, the dwell times in the metastable states also increase~\cite{Sett2024}, making the D$\to$B jumps more scarce. This is in contrast to the situation we observe if we set $g=0$ keeping all other couplings the same: Monte Carlo simulations show that the conditional cavity field relaxes fast to a state with a steady-state photon number $\varepsilon^2/[\kappa^2 + (\Delta \omega)^2]$---in line with Eq.~\eqref{eq:neoclassical}. At the same time, the upward jumps make about half of the total, as is expected from a coherent state maintained by the balance of drive and decay.

How is then the dearth of upward jumps to be explained when we keep the light-matter coupling on, and how is photon bunching operationally registered with reference to the familiar photon correlations of a thermal state~\cite{Carmichael1993QTII}? The answer once more involves the regression of fluctuations. The route to photon bunching in bistability---operationally unfolding in individual realizations---is through establishing a bright state of an ever increasing photon occupation whence a higher $p_{\rm REC}(t)=2\kappa \langle n (t) \rangle_{\rm REC} \Delta t$. Even a vanishing number of upward jumps is enough to trigger the switch the bright state, from where the overwhelming majority of photon emissions will originate and registered as photon ``clicks'' from downward jumps. In that sense, the metastable bright state may be viewed as a very long regression (the state is long-lived with respect to $\kappa^{-1}$) of an initial fluctuation. For a cavity driven by thermal light ($\overline{n}\sim 1$), fluctuations typically regress within cavity lifetime---see Figs 8.6 and 8.7 of~\cite{Carmichael1993QTII}. As we have seen in Sec.~\ref{sec:mpJC}, single-jump fluctuations along a bistable trajectory in the vacuum (dim) state are also short-lived---they decay within $\kappa^{-1}$. The key difference is that a large number of them  are required to establish a large and long-lived ``fluctuation'': the bright mestastable state. Emissions within such a bundle are separated by time intervals on the order of $g^{-1} \ll {\kappa}^{-1}$, a timescale involving the coherent light-matter coupling. Hence, a more subtle manifestation of coherence for ``zero system size'' underlies the D$\to$B switching events, than the coherent localization we encounter in the B$\to$D jumps. Such background of a coherent continuous evolution (scaled by $g$) is of course absent in a cavity driven by thermal light. Therefore, the strong-coupling limit $n_{\rm scale}=[g/(2\kappa)]^2 \to \infty$ is associated with increasing photon bunching, in parallel with the scarcity of upward jumps. The process of {\it quantum activation} discussed in~\cite{Dykman1979, Dykman2009, Peano2010, Dykman2012} acquires thus its operational interpretation. 

There is a more subtle connection between photon bunching and bistable switching, expressed via an increased stability of conditional states and their superpositions. Figure~\ref{fig:FIG9} shows two jumps to and from the bright state for a linear scan of the drive amplitude with rate equal to $\approx 0.016\kappa$ per $\kappa^{-1}$, occurring when the dim state dominates in the corresponding steady-state {\it quasi}probability distribution [frame (d)], to be followed by an eventual D$\to$B jump in line with a dominant peak of the bright state in the steady-state $Q$ function. The upward jumps along the trajectory still make less than $1\%$ of the total, with highly bunched bundles concentrated around the two D$\to$B jumps. Despite the varying drive amplitude and the improbable occurrence of the bright state with reference to steady-state solution of the ME [the ``shrunk'' quantum vs. semiclassical bistability region for the same value of $g/\kappa$ is depicted in Fig. 2(a) of~\cite{Carmichael2015}], the invertible decay of a coherent-state superposition remains in place, with the two state amplitudes to be read from Fig.~\ref{fig:FIG9}(c). The corresponding null-measurement record obtained for an empty cavity is still in good agreement with the photoelectron counting record unraveling ME~\eqref{eq:ME}, predicting a localization time $2\kappa\Delta t_{\rm end}\approx 0.216$; equation~\eqref{eq:jumptime} yields the lower bound of $2\kappa\Delta t_{\rm end,\,min}\approx 0.146$. The conditional cavity tomograms testify to a pure state comprising a coherent state superposition.

The quantum jumps recorded along individual trajectories unraveling ME~\eqref{eq:ME} are contextual, depending on the measurement apparatus encountered by the output field. In that respect, a heterodyne measurement is a notable alternative to direct photodetection, where the conditional photon number is correlated with shot noise generated over the recent past. Such noise is being filtered through by the dynamical response of the system---the JC Hamiltonian---when a classical stochastic current is produced as measurement data. The normalized conditional state $|\tilde{\psi}^{\prime}_{\rm REC}\rangle$ obeys a nonlinear stochastic differential equation. In particular, it solves the following {\it stochastic Schr\"{o}dinger equation} (SSE) conditioned on heterodyne-current records~\cite{CarmichaelBook2}:

\begin{widetext}
\begin{equation}\label{eq:HetSSEJC}
\begin{aligned}
d|\tilde{\psi}^{\prime}_{\rm REC}\rangle=&\Bigg\{\frac{1}{i\hbar}\left[H-\tfrac{1}{2}\langle \tilde{\psi}^{\prime}_{\rm REC}|(H-H^{\dagger}+i\hbar J_{\rightarrow}^{\dagger}J_{\rightarrow})|\tilde{\psi}^{\prime}_{\rm REC}\rangle \right]dt -\tfrac{1}{2}\langle \tilde{\psi}^{\prime}_{\rm REC}|J_{\rightarrow}^{\dagger}|\tilde{\psi}^{\prime}_{\rm REC}\rangle \langle \tilde{\psi}^{\prime}_{\rm REC}|J_{\rightarrow}|\tilde{\psi}^{\prime}_{\rm REC}\rangle\,dt \\
& +\langle \tilde{\psi}^{\prime}_{\rm REC}|J_{\rightarrow}^{\dagger}|\tilde{\psi}^{\prime}_{\rm REC}\rangle J_{\rightarrow}\,dt +dZ \left(J_{\rightarrow}-\langle \tilde{\psi}^{\prime}_{\rm REC}|J_{\rightarrow}|\tilde{\psi}^{\prime}_{\rm REC}\rangle \right) \Bigg\}|\tilde{\psi}^{\prime}_{\rm REC}\rangle,
\end{aligned}
\end{equation}
\end{widetext}
where $H=H_{\rm JCD}-i\tfrac{1}{2}\hbar J_{\rightarrow}^{\dagger}J_{\rightarrow}$ is the familiar non-Hermitian JC Hamiltonian we used in the Monte Carlo algorithm of Appendix~\ref{sec:MC}, and $J_{\rightarrow}=\sqrt{2\kappa}\,a$ is the photon jump operator. Shot noise is modelled by $dZ=(dW_x+i\,dW_y)/\sqrt{2}$, a complex-valued Wiener increment with covariances $\overline{dZ dZ}=\overline{dZ^{*}dZ^{*}}=0$ and $\overline{dZ^{*}dZ}=dt$. The statistically independent Wiener increments $dW_{x}$ and $dW_{y}$, corresponding to the two quadratures of heterodyne detection, have covariances $\overline{dW_{x}dW_{x}}=\overline{dW_{y}dW_{y}}=dt$ and $\overline{dW_{x}dW_{y}}=0$. The label ${\rm REC}$ in~\eqref{eq:HetSSEJC} qualifies heterodyne detection. Equation~\eqref{eq:HetSSEJC} is numerically solved using a Cash--Karp Runge--Kutta algorithm with adaptive time steps for its deterministic part, and first--order Euler integration for the stochastic part~\cite{Schack1997}.
\begin{figure*}
\centering
\includegraphics[width=\textwidth]{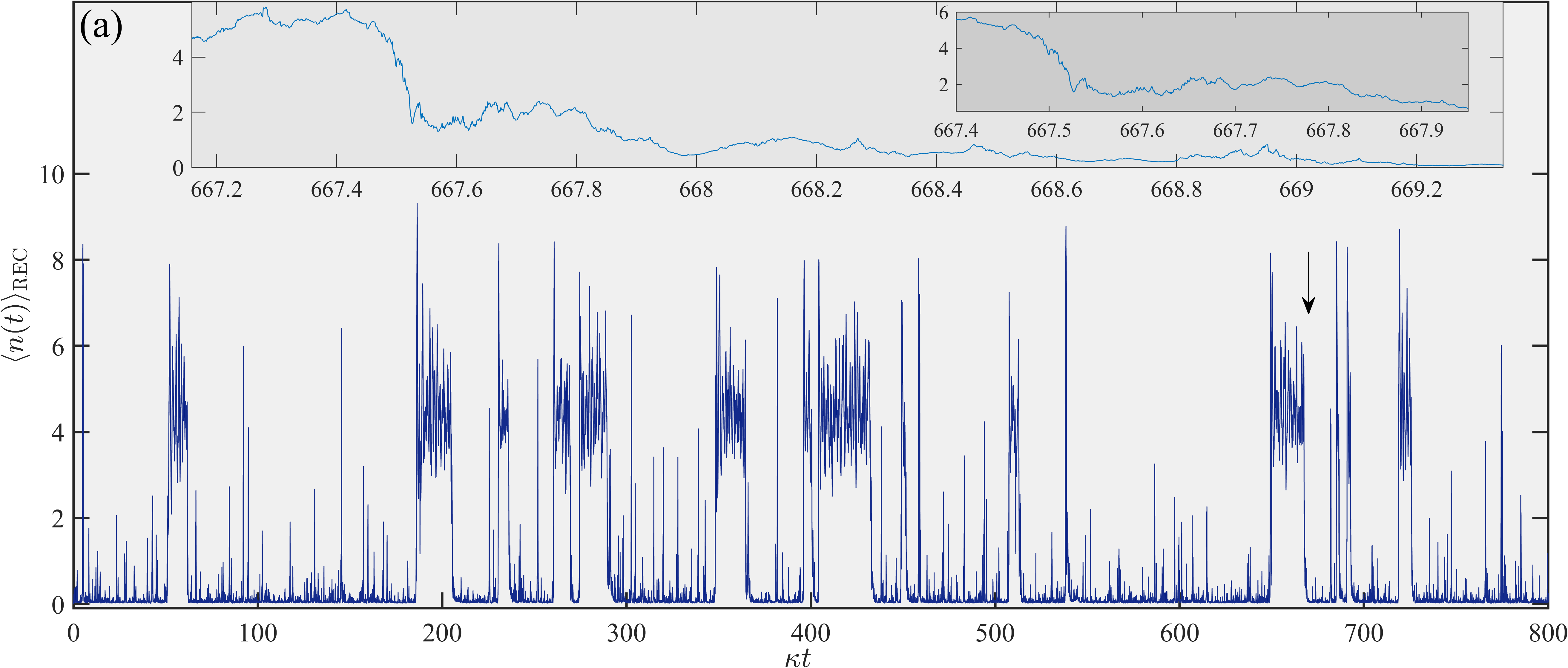}
\includegraphics[width=\textwidth]{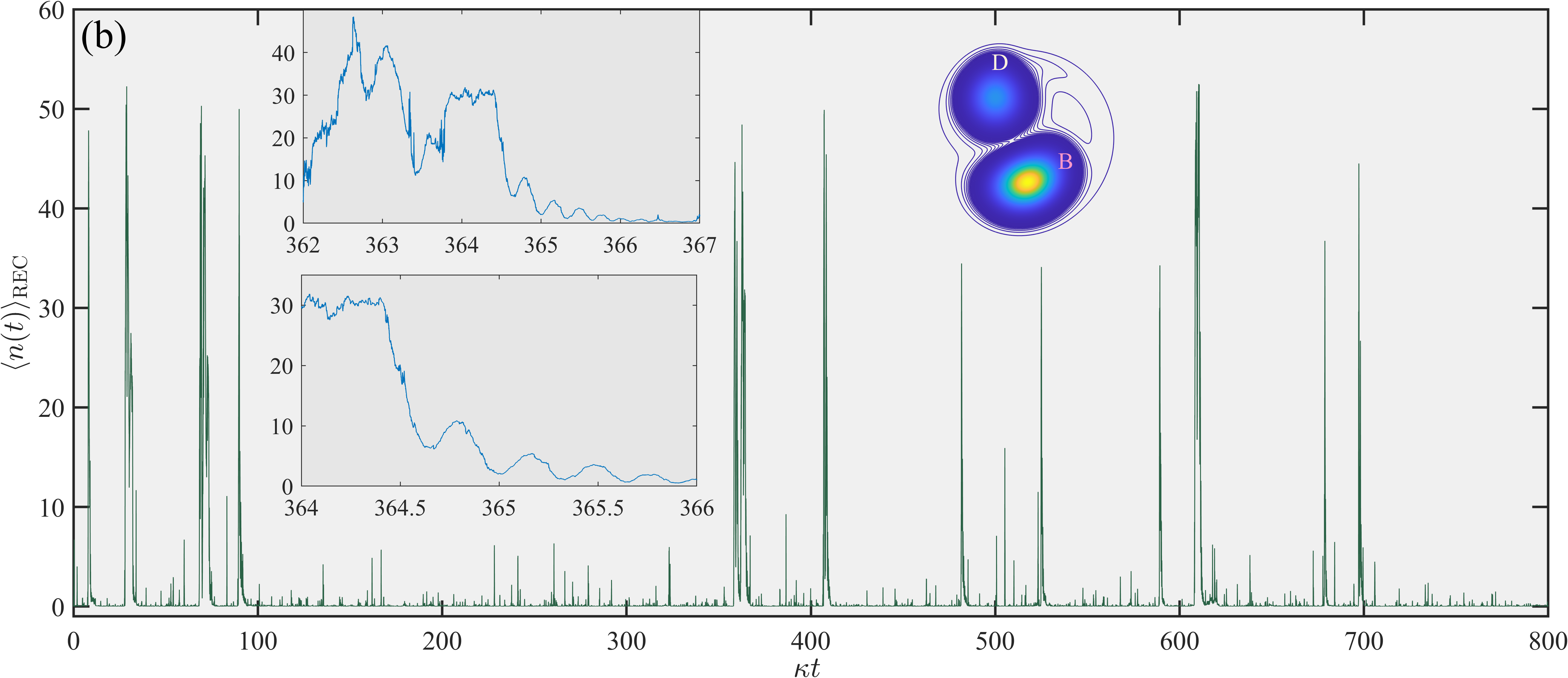}
\caption{{\it Approaching the strong-coupling ``thermodynamic limit'' with heterodyne detection.} {\bf (a)} Conditional average photon number $\langle n(t) \rangle^{\rm JC}_{\rm REC,\, heterodyne}$ along a sample trajectory unraveling ME~\eqref{eq:ME} over $800$ cavity lifetimes using a heterodyne detection scheme. The larger inset on top focuses on the B$\to$D jump indicated by the arrow, zoomed in even further in the smaller inset. The operating parameters read: $g/\kappa=25$, $\varepsilon/\kappa=5.3$, and $\Delta\omega/\kappa=-8$, while the Fock-state basis is truncated at $L_{\rm max}=20$. {\bf (b)} Same as in (a) but with operating parameters: $g/\kappa=50$, $\varepsilon/\kappa=13.5$, and $\Delta\omega/\kappa=-8$. The two insets on the left focus on the regression of a large-amplitude fluctuation of about $10$ cavity lifetimes, while the inset on the right depicts a schematic contour of $Q_{\rm ss}(x+iy)$. D and B denote the dim and bright states, respectively. The Fock-state basis is truncated at $L_{\rm max}=20$ and $60$, in (a) and (b), respectively. The time step is taken $\kappa \Delta t=0.001$ and $0.0005$ in (a) and (b), respectively, much smaller than the fastest timescale~$\sim \kappa/g$ associated with the approach to the ``thermodynamic limit''.}
\label{fig:FIG10}
\end{figure*}

Figure~\ref{fig:FIG10}(a) illustrates a heterodyne detection record made while operating with the same conditions as in Fig.~\ref{fig:FIG2}. The average switching rate in the conditional photon number is similar to direct photodetection, but the lifetime of the bright state has been visibly reduced. With higher values of $g/\kappa$ and for drive parameters producing steady-state bimodality, as in Fig.~\ref{fig:FIG3}, we find that the lifetime of the bright state is further reduced, in favour of a larger fluctuation accompanying every D$\to$B jump. This trend is also followed when the bright state dominates in the steady-state response, as shown in Fig.~\ref{fig:FIG10}(b). The decreased lifetime of the bright state in parallel with the high-amplitude fluctuations in a D$\to$B jump form a type of ``uncertainty relation'' in the heterodyne-detection unraveling of the ``thermodynamic limit'', working in a `complementary' direction to direct photodetection. In the latter, we saw earlier that the bright-state lifetime increases as $g/\kappa \to \infty$, while there is a pronounced asymmetry established between upward and downward jumps. In other words, in direct photodetection, the two quantities observed to follow an operational `uncertainty relation' are the lifetime of the bright state---translated to a long series of recorded photon ``clicks''---on the one hand, and the ratio of upward to total jumps on the other. To make the connection between the two unravelings, we note that in heterodyne detection the measured signal is a photocurrent whose `deterministic' part doesn't involve the conditional photon number, but $\langle \tilde{\psi}^{\prime}_{\rm REC}|J_{\rightarrow}^{\dagger}|\tilde{\psi}^{\prime}_{\rm REC}\rangle$ instead---see Sec.~\ref{subsec:MQ} and Ref.~\cite{CarmichaelBook2}.   

The two insets of Figs.~\ref{fig:FIG10}(a, b) show that a B$\to$D jump is completed roughly within an average photon lifetime, while the full jump is completed in two stages, similar to the scattering record collected under direct detection. Correlation with the injected shot noise, however, doesn't allow for a coherent-state superposition to coherently decay as in the continuous monitoring of direct photon counting, nor does it permit an invertible evolution. Is this all that can be said about the use of complementary unravelings in amplitude bistability? Key to the answer in the negative is the time separation of two detection schemes: direct photodetection is to be followed by heterodyne measurement on the field of a cavity where light-matter coupling is switched off, once a coherent-state superposition has been formed. This brings us to the penultimate section of our paper, where we will see how can heterodyne detection be linked to the detection of a coherent-state superposition. 

\begin{figure*}
\centering
\includegraphics[width=\textwidth]{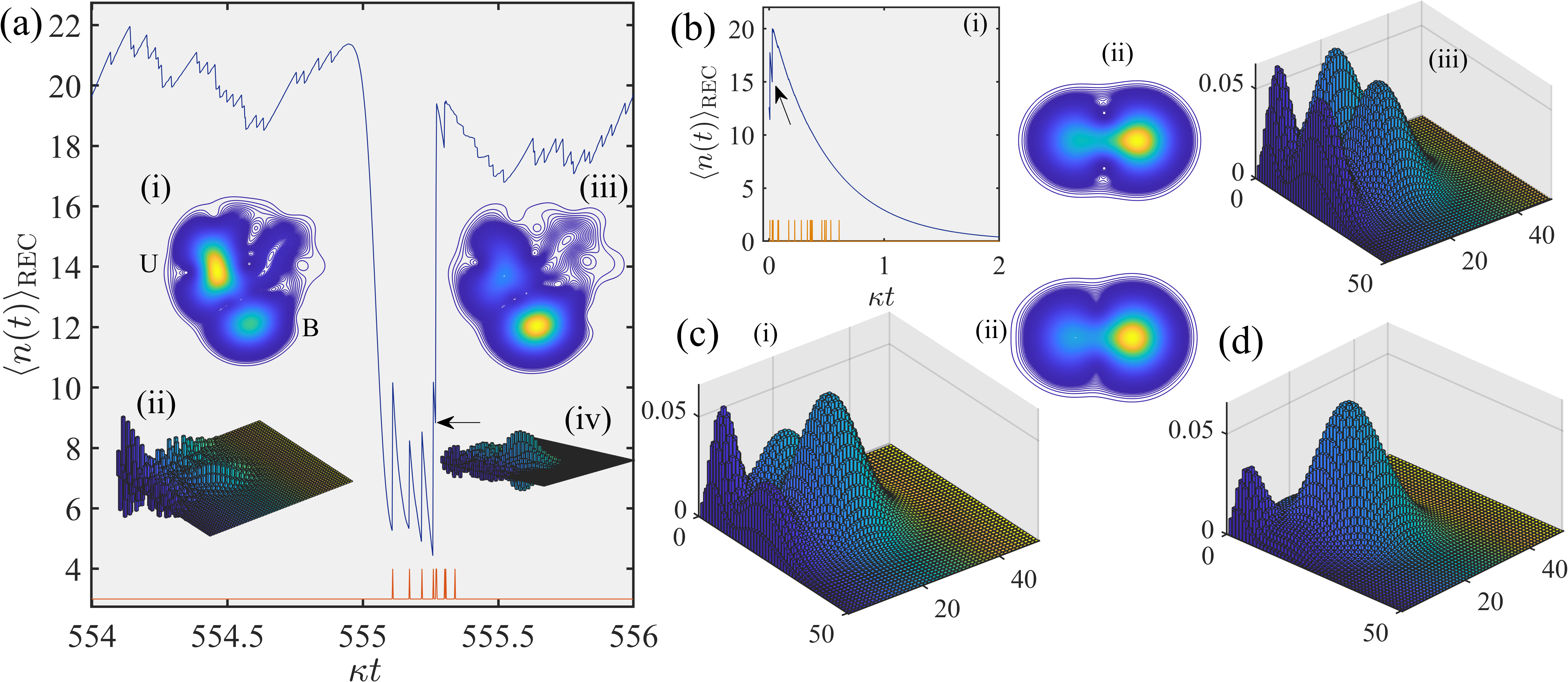}
\caption{{\it Detector inefficiency and thermal photons.} {\bf (a)} Focus on an incomplete B$\to$D jump along  $\langle n(t)\rangle^{\rm JCD}_{\rm REC}$ for $2$ cavity lifetimes. Insets (i-iv): Schematic contour plots of the conditional distribution $Q_{\rm REC}(x+iy;t_k)$ and real parts of $(\rho_{\rm cav;\,REC}(t_k))_{mn}$ corresponding to the photon averages $\langle n(t_k)\rangle^{\rm JCD}_{\rm REC}=8.82\, {\rm (i, ii)}$ and $16.49\, {\rm (iii, iv)}$ in the course of the localization to the bright state marked by the left pointing arrow. {\bf (b)} (i) Sample trajectory unraveling ME~\eqref{eq:MEdm}, depicting the conditioned photon number for a localization to the high-amplitude state $\alpha_1$ with initial state~\eqref{eq:inputStatemain}. The two amplitudes $\alpha_1, \alpha_2$ are taken real and positive, having the same magnitude as the bright and unstable states, respectively, of the localization shown in (a): $\alpha_1=|\alpha_1|=4.64$ and $\alpha_2=|\alpha_2|=1.97$. The schematic contour (ii) and barplot (iii) depict $Q_{\rm REC}(x+iy;t_1)$ and $(\rho_{\rm REC}(t_1))_{mn}$ [the right peak is centred at $\alpha_1$], respectively, corresponding to the conditional average photon number $\langle n(t_1)\rangle^{\rm JCD}_{\rm REC}=15.00$, indicated by the arrow in the trajectory of (i). {\bf (c)} Barplot of $(\rho_{\rm REC}(t_1^{\prime}))_{mn}$ (i) and schematic contour of $Q_{\rm REC}(x+iy;t_1^{\prime})$ (ii) corresponding to the conditional average photon number $\langle n(t_1^{\prime})\rangle^{\rm JCD}_{\rm REC}=14.81$ along a similar localization trajectory to that of (b), but for a detector efficiency $\eta=0.5$. {\bf (d)} Barplot of $(\rho_{\rm REC}(t_1^{\prime\prime}))_{mn}$ corresponding to the conditional average photon number $\langle n(t_1^{\prime\prime})\rangle^{\rm JCD}_{\rm REC}=15.55$ along a similar localization trajectory to that of (b), but for a detector efficiency $\eta=0.5$ and a thermally excited reservoir with $\overline{n}=1$. The operating parameters for the trajectory in (a) read: $g/\kappa=50$, $\varepsilon/\kappa=11.6$ and $\Delta\omega/\kappa=-8$. The Fock-state basis is truncated at $L_{\rm max}=50$.}
\label{fig:FIG11}
\end{figure*}

\section{Experimentally assessing a coherent-state localization}
\label{sec:exp}

\subsection{Sources of incoherence}
\label{subsec:sourcesincM}

Before we move to complementary unraveling methods, let us briefly stay with direct photodetection and revert to its link with the empty-cavity configuration, but in a situation where the ME of the damped harmonic oscillator {\it cannot} be unraveled into pure states. The next figure we meet here is devoted to an `inverse jump', showing a trajectory where the upward jumps precipitate the localization to the bright state (see once more the comment on the competition between the number of jumps and the coherent evolution in Appendix~\ref{subsubsec:dampedcs}). In particular, Fig.~\ref{fig:FIG11}(a) shows an incipient B$\to$D jump interrupted by a series of upward jumps managing to revert the coherent localization to the bright state. The conditional cavity distributions taken during the (second) localization process detail the probability transfer from the unstable to the bright state, while the barplots of the conditional density matrix clearly show off-diagonal peaks at both instances. We point out that this process is different to the mechanism underlying the occurrence of D$\to$B jumps, where there is no coherent localization but only single states when climbing up the excitation path. Upward jumps are present in both cases, but the underlying coherent null-measurement evolution which is absent in the D$\to$B switching instances between established metastable states; in the trajectory of Fig.~\ref{fig:FIG11}(a), the dim state is not established.

Further along, two sources of incoherence are taken into account, in line with our discussion in Appendix~\ref{subsec:sourcesinc}. An example of a trajectory unraveling ME~\eqref{eq:MEdm} and initial condition~\eqref{eq:inputStatemain}, localized to the high-amplitude state, is shown in Fig.~\ref{fig:FIG11}(b). The conditional $Q$ function and cavity density matrix set the reference for perfect detection with efficiency $\eta=1$, and a reservoir in the vacuum state with thermal occupation $\overline{n}=0$. 

Figure~\ref{fig:FIG11}(c) illustrates the impact of a detector with efficiency $\eta=0.5$ intercepting the output beam in a quantum trajectory generated for otherwise identical conditions to frame (b). Off-diagonal distributions in the cavity density matrix, corresponding to a photon-number average very close to that of frame (b), are reduced in amplitude yet still present. When, besides imperfect detection, the cavity mode is driven by thermal light, the coherences are gone [Fig.~\ref{fig:FIG11}(d)]. We infer that localization, as a coherent process, is more prone to thermal excitation than to a limited detector efficiency. Therefore, it is crucial for experiments at microwave frequencies to operate at very low temperatures in order to attribute the jumps to a coherent-state localization record. 

\subsection{Measuring a $Q$-distribution through heterodyne detection}
\label{subsec:MQ}

We will now present a method for determining the conditional distribution of the intracavity field in the course of bistable switching, which is intimately tied to making a choice between alternative ``meter'' states in a quantum measurement---what localization amounts to via the decoherence of a state superposition. To develop the required formalism, we draw on Sec. 18.3.2. of~\cite{CarmichaelBook2}, and examine a complementary unraveling of the ME~\eqref{eq:MEdm}---see also~\cite{SGComment1994, Carmichael1994, Carmichael1999}. We imagine that, at a certain time $t=t_{\rm off}$ during a single run of an experiment measuring the output channel of ME~\eqref{eq:ME}, we turn off the drive $\varepsilon$ and the light-matter coupling strength $g$. The cavity field is then allowed to freely decay, and the record keeping is switched to heterodyne detection. We restart the clock and consider once again an empty cavity prepared in the pure state $|\psi(0)\rangle$, the same as the conditional state of the bistable system at $t=t_{\rm off}$: $|\psi(0)\rangle=|\psi^{\rm JCD}_{\rm REC}(t_{\rm off})\rangle$.  We will use a local oscillator of amplitude $|\mathcal{E}_{\rm lo}|$ that is far detuned with respect to the cavity resonance frequency. The mismatch $\Delta \omega$ is much larger than $\kappa$--the bandwidth of the source-field fluctuations. 

The linear SSE governing the free decay of the un-normalized quantum state of the intracavity field conditioned on temporally mode-matched heterodyne detection (one where the amplitude of the local oscillator decays as $|\mathcal{E}_{\rm lo}|e^{-\kappa t}$) reads
\begin{equation}\label{eq:SSE}
d|\tilde{\psi}_{\rm REC}\rangle=\left(-\kappa a^{\dagger}a\,dt +\sqrt{2\kappa}\, a\,d\xi\right)|\tilde{\psi}_{\rm REC}\rangle,
\end{equation}
where
\begin{equation}
d\xi \equiv e^{\kappa t}(Ge|\mathcal{E}_{\rm lo}|)^{-1}d\tilde{q}=\sqrt{2\kappa} \langle a^{\dagger} \rangle_{\rm REC}dt +dZ.
\end{equation}
In the above relation, $d\tilde{q}=e^{i\Delta \omega t}dq$ is the slowly-varying incremental charge deposited in the detector circuit in the interval $t$ to $t+dt$, $e$ is the electronic charge and $G$ is the detector gain. The label ${\rm REC}$ in~\eqref{eq:SSE} qualifies heterodyne detection. 

Our aim is to determine the distribution $P(\tilde{Q}, \tilde{Q}^{*},t)$ of the cumulative complex charge
\begin{equation}
\tilde{Q} \equiv \sqrt{2\kappa}\,(G e |\mathcal{E}|_{\rm lo})^{-1}\int_{0}^{t}d\tilde{q}=\sqrt{2\kappa}\int_{0}^{t}e^{-\kappa t^{\prime}}d\xi^{\prime}.
\end{equation}
The incremental cumulative charge deposited in the detector circuit in time step $dt$ is
\begin{equation}\label{eq:dQ}
\begin{aligned}
d\tilde{Q}&=\sqrt{2\kappa}\,e^{-\kappa t}\,d\xi \\
&=e^{\kappa t}\langle a^{\dagger}(t)\rangle_{\rm REC}(2\kappa e^{-2\kappa t}\,dt) + \sqrt{2\kappa}\, e^{-\kappa t}dZ. 
\end{aligned}
\end{equation} 
We need to first solve Eq.~\eqref{eq:SSE} to determine the conditional average $\langle a^{\dagger}(t)\rangle_{\rm REC}$, which is to be substituted in Eq.~\eqref{eq:dQ} to formulate a stochastic differential equation for the cumulative charge. We begin by setting 
\begin{equation}
|\tilde{\psi}_{\rm REC}\rangle=e^{-\kappa a^{\dagger}a\,t}|\chi\rangle,
\end{equation} 
which transforms Eq.~\eqref{eq:SSE} to
\begin{equation}
\begin{aligned}
d|\chi\rangle&=\sqrt{2\kappa}\,e^{\kappa a^{\dagger}a t}\, a\, e^{-\kappa a^{\dagger}a t}\,d\xi|\chi\rangle\\
&=\sqrt{2\kappa}\,e^{-\kappa t}\,a\,d\xi\,|\chi\rangle\\
&=a\,d\tilde{Q}|\chi\rangle,
\end{aligned}
\end{equation}
with solution $|\chi\rangle=e^{\tilde{Q}a}|\chi(0)\rangle$. Hence, the solution to Eq.~\eqref{eq:SSE} is obtained in terms of the initial pure state of interest $|\psi(0)\rangle$ as
\begin{equation}\label{eq:psiRECH}
|\tilde{\psi}_{\rm REC}(t)\rangle=e^{-\kappa a^{\dagger}a t} e^{\tilde{Q}a}|\psi(0)\rangle.
\end{equation}
This allows us to evaluate the conditional expectation
\begin{equation}
\begin{aligned}
\langle a^{\dagger}(t)\rangle_{\rm REC}&=\frac{\langle \psi(0)| e^{\tilde{Q}^{*}a^{\dagger}} e^{-\kappa a^{\dagger}a t} a^{\dagger} e^{-\kappa a^{\dagger}a t} e^{\tilde{Q}a}|\psi(0)\rangle}{\langle \psi(0)| e^{\tilde{Q}^{*}a^{\dagger}} e^{-\kappa a^{\dagger}a t} e^{-\kappa a^{\dagger}a t} e^{\tilde{Q}a}|\psi(0)\rangle}\\
&=e^{-\kappa t} \frac{\partial}{\partial \tilde{Q}^{*}}\ln\left[\langle \psi(0)| e^{\tilde{Q}^{*}a^{\dagger}} e^{-2\kappa a^{\dagger}a t} e^{\tilde{Q}a}|\psi(0)\rangle \right].
\end{aligned}
\end{equation}
Substituting this result into Eq.~\eqref{eq:dQ}, we find that the cumulative charge $\tilde{Q}$ deposited in the heterodyne detector satisfies the stochastic differential equation
\begin{equation}\label{eq:SSEdQ}
d\tilde{Q}=-\frac{\partial}{\partial \tilde{Q}^{*}} V(\tilde{Q}, \tilde{Q}^{*},t) (2\kappa e^{-2\kappa t}\,dt) + \sqrt{2\kappa}\,e^{-\kappa t}\, dZ,
\end{equation}
where in anticipation of a ``drift'' term in a Fokker-Planck equation, we have introduced the time-dependent potential
\begin{equation}
V(\tilde{Q}, \tilde{Q}^{*},t)=-\ln\left[\langle \psi(0)| e^{\tilde{Q}^{*}a^{\dagger}} e^{-2\kappa a^{\dagger}a t} e^{\tilde{Q}a}|\psi(0)\rangle\right],
\end{equation} 
explicitly depending on the initial state. We make the variable change
\begin{equation}
\nu = 1-e^{-2\kappa t},
\end{equation}
to arrive at the simpler equation for the deposited charge:
\begin{equation}
d\tilde{Q}=-\frac{\partial}{\partial \tilde{Q}^{*}}V(\tilde{Q}, \tilde{Q}^{*},\nu)\,d\nu + d\zeta,
\end{equation}
where
\begin{equation}
V(\tilde{Q}, \tilde{Q}^{*},\nu)=-\ln\left[\langle \psi(0)| e^{\tilde{Q}^{*}a^{\dagger}} (1-\nu)^{a^{\dagger}a} e^{\tilde{Q}a}|\psi(0)\rangle\right],
\end{equation}
and $d\zeta$ is a Wiener increment with covariances $\overline{d\zeta d\zeta}=\overline{d\zeta^{*}d\zeta^{*}}=0$ and $\overline{d\zeta^{*}d\zeta}=d\nu$. It follows that the probability distribution over the cumulative complex charge $P^{\prime}(\tilde{Q}, \tilde{Q}^{*},\nu) \equiv P(\tilde{Q}, \tilde{Q}^{*},t)$ obeys the Fokker-Planck equation (FPE)
\begin{equation}
\begin{aligned}
\frac{\partial P^{\prime}}{\partial \nu}=&\Bigg\{\frac{\partial}{\partial \tilde{Q}}\left[\frac{\partial}{\partial \tilde{Q}^{*}} V(\tilde{Q}, \tilde{Q}^{*},\nu)\right]
\\&+ \frac{\partial}{\partial \tilde{Q}^{*}}\left[\frac{\partial}{\partial \tilde{Q}} V(\tilde{Q}, \tilde{Q}^{*},\nu)\right]+\frac{\partial^2}{\partial \tilde{Q}\partial \tilde{Q}^{*}}\Bigg\}P^{\prime}.
\end{aligned}
\end{equation} 

The above equation is satisfied by the distribution
\begin{equation}
\begin{aligned}
P^{\prime}(\tilde{Q}, \tilde{Q}^{*},\nu)&=\frac{1}{\pi \nu}e^{-|\tilde{Q}|^2/\nu}\,e^{-V(\tilde{Q}, \tilde{Q}^{*},\nu)}\\
&=\frac{1}{\pi \nu}e^{-|\tilde{Q}|^2/\nu}\,\langle \psi(0)| e^{\tilde{Q}^{*}a^{\dagger}} (1-\nu)^{a^{\dagger}a} e^{\tilde{Q}a}|\psi(0)\rangle.
\end{aligned}
\end{equation}
We seek the solution of the FPE in the steady state for $t\to\infty$ (or $\nu \to 1$). The first term in the Fock-state resolution of $(1-\nu)^{a^{\dagger}a}$ is $(1-\nu)^0 \,|0\rangle \langle 0|=|0\rangle \langle 0|$, while the remaining terms involve higher powers of $1-\nu$. Setting $\nu=1$, all the higher-power terms disappear, and we can write
\begin{equation}
\begin{aligned}
P(\tilde{Q},\tilde{Q}^{*},t\to\infty)&=\frac{1}{\pi} \langle \psi(0)|e^{\tilde{Q}^{*}a^{\dagger}}|0\rangle \langle 0|e^{\tilde{Q}a}|\psi(0)\rangle\\
&=\frac{1}{\pi}\langle \psi(0)|\tilde{Q}^{*}\rangle \langle \tilde{Q}^{*}|\psi(0)\rangle,
\end{aligned}
\end{equation}
where 
\begin{equation}
|\tilde{Q}^{*}\rangle=e^{-\frac{1}{2}|\tilde{Q}|^2}e^{\tilde{Q}^{*}a^{\dagger}}|0\rangle
\end{equation}
is the coherent state with complex amplitude $\tilde{Q}^{*}$. From the definition of the $Q$ function, we can then identify
\begin{equation}
P(\tilde{Q},\tilde{Q}^{*},t\to\infty)=Q_{0}(\tilde{Q}^{*},\tilde{Q}),
\end{equation}
where the right-hand side represents the $Q$ distribution of the initial pure state of the intracavity field, with density operator $\rho(0)=|\psi(0)\rangle \langle \psi(0)|$. We arrived at the result $Q_{0}(\tilde{Q}^{*},\tilde{Q})$ instead of $Q_{0}(\tilde{Q},\tilde{Q}^{*})$ because the equation for the incremental cumulative charge $d\tilde{Q}$ involves the conditional average $\langle a^{\dagger}(t)\rangle_{\rm REC}$ instead of $\langle a(t)\rangle_{\rm REC}$.

\subsection{Emergence of a double-well ``potential''}
\label{subsec:DW}

Let us now select a coherent-state superposition as the initial state,
\begin{equation}
|\psi(0)\rangle=\frac{1}{\sqrt{2}}\frac{|\alpha_1\rangle + |\alpha_2\rangle}{1+{\rm Re}\langle \alpha_1|\alpha_2\rangle}\approx \frac{1}{\sqrt{2}}(|\alpha_1\rangle + |\alpha_2\rangle),
\end{equation}
assuming $|\alpha_1-\alpha_2|^2 \gg 1$. Then, the solution to the SSE~\eqref{eq:SSE} is obtained as
\begin{equation}\label{eq:SSEsolCoh}
\begin{aligned}
|\tilde{\psi}_{\rm REC}(t)\rangle=&\frac{1}{\sqrt{2}}\Big\{ \exp\left[-\tfrac{1}{2}|\alpha_1|^2(1-e^{-2\kappa t}) \right]e^{\tilde{Q}\alpha_1}|\alpha_1 e^{-\kappa t}\rangle\\
& + \exp\left[-\tfrac{1}{2}|\alpha_2|^2(1-e^{-2\kappa t}) \right]e^{\tilde{Q}\alpha_2}|\alpha_2 e^{-\kappa t}\rangle \Big\},
\end{aligned}
\end{equation}
while the time-dependent ``potential'' to be substituted in Eq.~\eqref{eq:SSEdQ} is evaluated as
\begin{equation}
\begin{aligned}
V(\tilde{Q}, \tilde{Q}^{*},t)=&-\ln\Big\{\tfrac{1}{2}\exp\left[-|\alpha_1|^2(1-e^{-2\kappa t}) \right]e^{\tilde{Q}\alpha_1 + \tilde{Q}^{*}\alpha_1^*}\\
& + \tfrac{1}{2}\exp\left[-|\alpha_2|^2(1-e^{-2\kappa t}) \right]e^{\tilde{Q}\alpha_2 + \tilde{Q}^{*}\alpha_2^*} \Big\},
\end{aligned}
\end{equation}
where we have neglected terms $\sim |\langle \alpha_1 |\alpha_2 \rangle|$. In the limit $t\to \infty$, the potential acquires the form
\begin{equation}
V(\tilde{Q}, \tilde{Q}^{*},t\to \infty)=-\ln\left(\tfrac{1}{2}\sum_{j=1}^{2}e^{-|\alpha_j-\tilde{Q}^{*}|^2}e^{|\tilde{Q}|^2} \right).
\end{equation}
The above limit suggests the emergence of a double-well potential in the complex plane of the cumulative charge, with minima centred at the location of the two coherent-state amplitudes in the initial superposition. As we will find below, this instance points to the possibility of projective measurements. 
 
\subsection{Projective measurement and meter states}
\label{subsec:proj}

We will now see how coherent localization can be connected to a projective measurement. Let us consider the entangled initial state
\begin{equation}
|\phi(0)\rangle=\sum_{j=1}^{N}c_j |b_j \rangle |\psi_{j}(0)\rangle,
\end{equation}
where $|b_j\rangle$, $j=1, 2$ are two orthogonal eigenstates of an observable $\hat{B}$, and $|\psi_{j}(0)\rangle$ is a meter state, a component of the coherent-state superposition in the cavity mode that is correlated with $|b_{j}\rangle$. The cumulative charge distribution in the steady state reads
\begin{equation}\label{eq:pQent}
P(\tilde{Q},\tilde{Q}^{*},t\to\infty)=\sum_{j=1}^{N}|c_j|^2 Q_0^{(j)}(\tilde{Q}^{*},\tilde{Q}),
\end{equation}
where $Q_0^{(j)}(\tilde{Q}^{*},\tilde{Q})$ are localized distributions in phase space, representing the meter states $|\psi_{j}(0)\rangle$, $j=1, 2, \ldots N$. Each one is coordinated with a particular eigenstate of $\hat{B}$. The coefficient $|c_j|^2$ is the probability of obtaining a record coordinated with $|b_j\rangle$, as it happens in a projective measurement. 

Coherent states are ideally suited to serve as meter states, since they remain coherent in the course of the cavity decay; see Secs.~\ref{subsubsec:dampedA} and~\ref{subsec:DW}. Selecting $|\psi_{j}(0)\rangle=|\alpha_j\rangle$, we find the conditional wavefunction
\begin{equation}
|\tilde{\psi}_{\rm REC}(t)\rangle=\sum_{j=1}^{N}c_j  \exp\left[\lambda(|\alpha_j|^2/2;t) \right] e^{\tilde{Q}\alpha_j}|b_j\rangle |e^{-\kappa t}\alpha_{j}\rangle.
\end{equation}  
As we saw in Eq.~\eqref{eq:SSEsolCoh}, the term $\lambda(|\alpha_j|^2/2;t)\equiv\exp\left[-\frac{1}{2}|\alpha_j|^2(1-e^{-2\kappa t}) \right]$ is the probability density for a null measurement result, arising when we substitute $|\psi(0)\rangle$ with $|\alpha_j\rangle$ in Eq.~\eqref{eq:psiRECH}. In the steady state ($t \to \infty$), we obtain
\begin{equation}\label{eq:sumF}
|\tilde{\psi}_{\rm REC}(t\to \infty)\rangle=\sum_{j=1}^{N}c_j e^{-\frac{1}{2}|\alpha_j|^2}e^{\tilde{Q}\alpha_j}|b_j\rangle |0\rangle.
\end{equation}
Now the relative weighting between the different components of the superposition is given by the square of the new expansion coefficients:
\begin{equation}
\left|c_j e^{-\frac{1}{2}|\alpha_j|^2}e^{\tilde{Q}\alpha_j}\right|^2=|c_j|^2 e^{-|\alpha_j-\tilde{Q}^{*}|^2}e^{|\tilde{Q}|^2}. 
\end{equation}
We conclude that if the coherent states $\alpha_j$, $j=1, 2,\ldots N$ are macroscopically distinct (well localized in phase space), then from Eq.~\eqref{eq:pQent} the charge $\tilde{Q}^{*}$ will be closer to one particular amplitude $\alpha_k$, and sufficiently far from the rest. Consequently, only one coefficient ($j=k$) will be promoted in the sum~\eqref{eq:sumF} and, after normalization, the conditional state will be $|\alpha_k\rangle|0\rangle$, signifying the collapse of the wavefunction in the long-time limit of the trajectory. 

Coherent localization of meter states allows a different kind of measurement to the one based on the large change of the amplitude and phase of forced oscillations in bistability~\cite{Dykman2009}. Namely, a projective measurement on an observable, whose eigenstates are correlated with the bright and unstable states of bistability, is possible owing to the quantum coherence manifested {\it during} the switch. Every B$\to$D jump in the limit of  ``zero system size'' will produce a {\it quasi}coherent state superposition ``mid-flight''. On approaching the strong-coupling ``thermodynamic limit'' $n_{\rm scale}=[g/(2\kappa)]^2\to\infty$, these states will become macroscopically distinct and the projection into the corresponding eigenstate will be performed faster, according to the predictions of Eqs.~\eqref{eq:jumptime},~\eqref{eq:asympt}. To increase the number of terms in the sum~~\eqref{eq:sumF}, with different states participating in the various localization instances, we need to depart ME~\eqref{eq:ME} and its unraveling; a wider gamut of phase-space separated meter states is expected in systems with a larger number of coupled oscillators, exhibiting (amplitude and/or phase) multi-stability~\cite{Kilin1993, Delanty2011, Valagiannopoulos2022, Valagiannopoulos2025}.

\section{Spontaneous emission and structural instability of neoclassical states}
\label{sec:SE}

Atomic emission with rate $\gamma$ into modes other than the privileged cavity mode brings us to absorptive bistability~\cite{Bonifacio1978,Mandel1984,Rosenberger1991,Shirai2018,Mabuchi2018,Liu2022}, away from ``zero system size''~\cite{CarmichaelBook2, Carmichael2015}. For a single atom, it leads to a finite system size~\cite{Savage1988}, with the saturation photon number being $n^{\prime}_{\rm scale}=\gamma^2/(8g^2)$ -- one of a weak-coupling limit similar to the Kerr nonlinearity we visited in Sec.~\ref{sec:Kerr}. At the level of ensemble averages, we modify the RHS of~\eqref{eq:ME} to $\mathcal{L}_{\kappa}^{\rm JCD}\rho-(\gamma/2)(2\sigma_{-}\rho\sigma_{+}-\sigma_{+}\sigma_{-}\rho-\rho\sigma_{+}\sigma_{-})$. 

\begin{figure}
\centering
\includegraphics[width=0.475\textwidth]{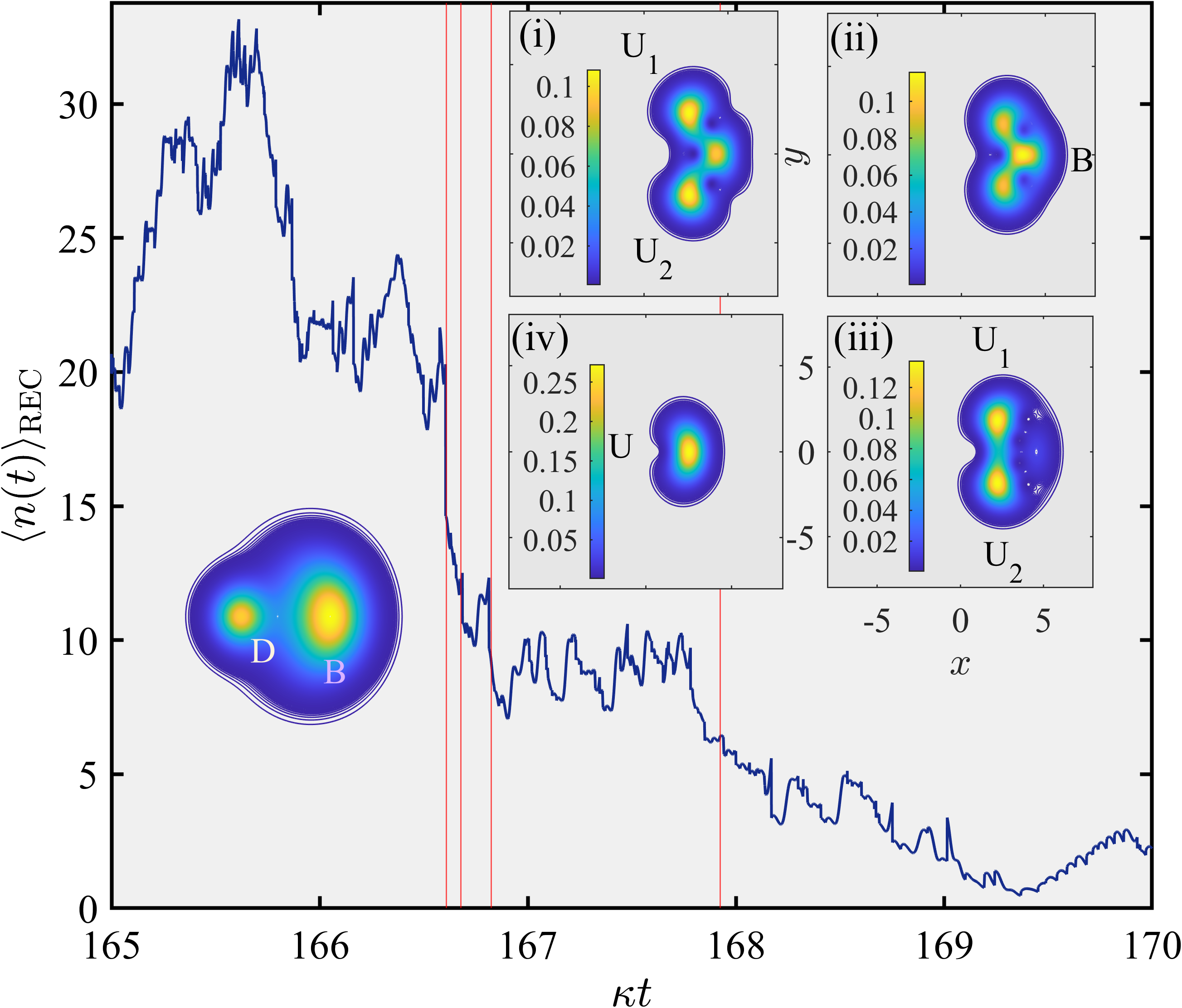}
\caption{{\it Anticipating phase bimodality: probability transfer in absorptive bistability.} Fragment of a sample realization depicting the conditioned photon number $\langle n(t)\rangle_{\rm REC}$ for $5$ cavity lifetimes, focusing on a B$\to$D jump. Insets (i--iv) on the right depict $Q_{\rm REC}(x+iy;t_k)$ where $\langle n(t_k)\rangle^{\rm JCD}_{\rm REC}=14.56, 12.18, 9.19$ and $6.40$, for $k=1 {\rm (i)},2 {\rm (ii)}, 3{\rm (iii)}, 4{\rm (iv)}$, respectively, marked by the four vertical lines. The schematic contour plot on the left shows $Q_{\rm ss}(x+iy)$. B, D, U, denote the bright, dim and unstable states, respectively. The operating parameters read: $g/\kappa=20$, $\varepsilon/\kappa=8.04$, $\gamma/\kappa=56$ and $\Delta\omega/\kappa=0$.}
\label{fig:FIG12}
\end{figure}

Two features in Fig.~\ref{fig:FIG12} are worth emphasizing: first, the switch takes much longer and is very frequently interrupted by spontaneous emission events. The saturable two-state absorber cycles through several atomic lifetimes, scattering many photons from the cavity in a time comparable to $\kappa^{-1}$~\cite{Kerckhoff11P}. Second, the unstable state splits into a pair of complex-conjugate states, through which population transfer from the bright state takes place. We remark that the conjugate pair appears in the conditioned phase-space distribution despite the fact that the drive amplitude $\varepsilon$ falls below its threshold value $g/2$ for the neoclassical bifurcation~\cite{Alsing1991, Carmichael2015}. Inset (iv) depicts an instance where the localization to the unstable state has been completed, following the merge of $U_1$ and $U_2$. The decay to the dim state takes another average photon lifetime, as was the case for the jumps in the ``zero system size'' limit.  

\section{Concluding remarks}

We have studied the mechanism underlying quantum jumps between {\it dynamical} states influenced in a fundamental way by the inputs and output of the JC interaction. We conclude that: {\bf (1)} while the process of state localization is inherent in the bistable switching to the vacuum, the ``zero system size'' limit of light-matter interaction allows for quantum jumps between macroscopic metastable states of light to be tracked by a null-measurement record registering the localization of a coherent-state superposition in an {\it empty cavity}; {\bf (2)} the said null-measurement evolution is contextual (particular to direct photodetection yet of a universal character), stable, continuous and deterministic; {\bf (3)} the duration of localization is a nonlinear function of the metastable state amplitudes, which are themselves nonlinear functions of the light-matter coupling strength, drive amplitude and photon loss rate. Coherent localization is faster towards higher-amplitude bistability; {\bf (4)} once the probability transfer between the bright and unstable states has been completed in the course of a localization instance, the unstable state decays to the vacuum typically within an average photon lifetime following a regression modelled by the neoclassical equations of motion; {\bf (5)} switching events from the dim to the bright metastable states do not involve a coherent-state superposition, but only a single squeezed state spiralling along the JC excitation path in phase space. Highly bunched photon bundles in the photon counting records condition the spiral ascent; {\bf (6)} photon counting records obtained from measuring the output field of the JC oscillator under amplitude bistability conditions evince a dearth of upward jumps with reference to the empty driven cavity of the same steady-state excitation, despite the fact that the autocorrelation function of the cavity field exceeds unity. The upwards jumps in question always accompany `intense' fluctuations from an established metastable state, including switching events; {\bf (7)} the formation of an unstable-state complex in phase space during a bright-to-dim switch heralds the interruption of coherent localization by a photon emission resolving the superposition. Such photon emissions prolong the downward switch, much like what is seen in absorptive optical bistability; {\bf (8)}   direct photodetection yields the shortest possible jumptime in the bistable switching. Interference with a coherent local-oscillator field in homodyne and heterodyne detection works against the coherent-state localization establishing Eq.~\eqref{eq:jumptime}. {\bf (9)} the coherent-state localization of JC bistability in its strong-coupling limit has distinct operational consequences. For example, it can be employed in conjunction with heterodyne detection to model a projective measurement, upon selecting the coherent states involved as meter states. 

We emphasize that, in our treatment, quantum fluctuations have been dealt with explicitly non perturbatively via stochastic processes in Hilbert space, correlated with detectable scattering events. Sample conditioned photoelectron counting records reveal details of bistable switching that remain hidden when we calculate the steady-state solution to the driven JC operator ME. Furthermore, coherence enables inversion of the conditioned intracavity excitation about the photon number of the initial superposition in a deterministic evolution for set amplitudes. A quantum-state tomogram~\cite{Welsch1999, Vogel2006, HarocheBook, Deleglise2008,Hofheinz2008,Eichler2012, Blais2021,Ahmed2021,Ahmed2021B} will directly reveal the coexistence of a metastable and an unstable state during the switch. Moreover, heterodyne detection on an empty cavity, prepared in a pure state obtained from a single experimental run, can visualize the conditioned $Q$-function at a given point during the switch~\cite{CarmichaelBook2}, as well as perform a projective measurement of an entangled observable eigenstate.

For every jump along an individual realization, the two states involved in the localization are squeezed and often not isolated in phase space, while their locations are subject to fluctuations visible {\it e.g.}, in phase space for different jumps. In fact, these fluctuations were explicitly taken into account in the derivation of Eq.~\eqref{eq:jumptime}, the lower time bound for coherent state localization. The bound can be evaluated with the bifurcating solutions of the MBEs as inputs; it naturally becomes more accurate as the ME solution converges to the mean-field predictions. Monte-Carlo simulations show that modelling the observed localization with a damped coherent-state superposition is in very good agreement with conditional records of state coherences.  

The attainment of the strong-coupling ``thermodynamic limit'' in quantum amplitude bistability is characterized by a growing scarcity of upward jumps and increased photon bunching, precipitating the switching to the high-amplitude state; in parallel, the dwell times in the two metastable states become longer. It is also associated with well localized conditional states (and their superpositions) which are rendered increasingly stable against quantum fluctuations, in contrast to the weak-coupling limit of a Kerr nonlinearity~\cite{WallsKerr,KheruntsyanKerr,Bishop2010,Miranowicz2013,Mavrogordatos2016}. For the switching to the vacuum in the same limit, the inversion of coherent localization asymptotically approaches time reversal of the null-measurement record about the localization midpoint.  

\appendix
\onecolumngrid 
\par\noindent\rule{\textwidth}{0.5pt}
\setcounter{equation}{0}

In the following two appendices, we provide a background to the theory of quantum trajectories necessary to understand the process of coherent localization. We derive analytical expressions for the null record density matrix and average photon number, and detail the numerical methods employed for the generation of simulated photoelectron counting records and the representation of the corresponding conditional states. We also formulate the argument used to derive a semiclassical expression for the  lower time bound predicted by the coherent localization process. Experimental limitations owing to sources of incoherence are briefly discussed.  

\section{Photoelectric counting records, Monte Carlo algorithm, conditional states and their representation}
\label{sec:MC}

The master equation governing the evolution of the system density matrix reads
\begin{equation}\label{eq:MElab}
\frac{d\rho}{dt}=\frac{1}{i\hbar}[H_{\rm JC}(t),\rho] + \kappa(2a\rho a^{\dagger}-a^{\dagger}a\rho-\rho a^{\dagger}a),
\end{equation}
where the JC Hamiltonian is explicitly time dependent owing to coherent driving, and is conveniently decomposed as:
\begin{equation}
H_{\rm JC}(t)=H_0 + H_g + H_D(t),
\end{equation}
where $H_0=\hbar\omega_0(\sigma_ {+}\sigma_{-}+a^{\dagger}a)\equiv \hbar \omega_0 N$ ($N$ is the operator of the total system excitation), $H_g=\hbar g (a^{\dagger}\sigma_{-}+a\sigma_{+})$ and $H_D(t)=i\hbar \varepsilon  (a^{\dagger}e^{-i\omega_d t}-a e^{i\omega_d t})$. The validity of the rotating wave approximation (RWA) is here ensured by taking $ \sqrt{n_{\rm ss}}(g/\omega_0)\lesssim 0.1$~\cite{Burgarth2024}, where $n_{\rm ss}$ is the steady-state cavity photon number. Transforming now to a frame rotating with the drive, $\rho \to U^{\dagger} \rho U$ with $U=\exp(i\omega_d N t)$, results in a ME of the form~\eqref{eq:MElab}, with 
\begin{equation}
H_{\rm JC}(t) \to U H_{\rm JC}(t) U^{\dagger}+i\hbar(dU/dt)U^{\dagger}=H_{\rm JCD}=-\hbar(\omega_d-\omega_0) (a^{\dagger}a + \sigma_{+}\sigma_{-})+i\hbar g(a^{\dagger}\sigma_{-}-a\sigma_{+})+i\hbar \varepsilon(a^{\dagger}-a),
\end{equation}
which is the time-independent Hamiltonian featuring in the ME~\eqref{eq:ME}.

We detail the numerical methods used to generate single realizations unraveling the master equations (MEs)
\begin{equation}\label{eq:MEJC}
\frac{d\rho}{dt}=\mathcal{L}_{\kappa}^{\rm JCD}\rho=\frac{1}{i\hbar}[H_{\rm JCD},\rho] + \kappa(2a\rho a^{\dagger}-a^{\dagger}a\rho-\rho a^{\dagger}a)
\end{equation}
and
\begin{equation}\label{eq:MEd}
\frac{d\rho}{dt}=\mathcal{L}\rho=\kappa(2a\rho a^{\dagger}-a^{\dagger}a\rho-\rho a^{\dagger}a).
\end{equation}
The unraveling of Eq.~\eqref{eq:MEJC} is used to generate sample trajectories evincing bistable switching in the driven dissipative JC model, while several numerically-generated trajectories unraveling Eq.~\eqref{eq:MEd} featured in the main text, to illustrate aspects of coherent localization.

Let us focus for simplicity on the ME~\eqref{eq:MEd}. In the Born-Markov approximation, the exclusive probability density for the realization of a particular photoelectron counting record, with precisely $n$ photons in the output field at times $t_1, t_2, \ldots, t_n$ and no photon at any other time in the interval from $0$ to $t$, is written as a trace over the system ($S$) degrees of freedom~\cite{Carmichael1993QTI, CarmichaelBook2},
\begin{equation}
{\rm tr}_S[\overline{\rho}_{\rm REC}(t)]={\rm tr}_S [e^{(\mathcal{L}-\mathcal{S})(t-t_n)}\mathcal{S}e^{(\mathcal{L}-\mathcal{S})(t_n-t_{n-1})}\ldots \mathcal{S}e^{(\mathcal{L}-\mathcal{S})t_1}\rho(0)],
\end{equation}
where $\overline{\rho}_{\rm REC}(t)$ is the un-normalized conditional state (indicated by the overbar), and $\rho(0)$ is the initial system state. We have made the decomposition
\begin{equation}
\mathcal{L}=\mathcal{L}-\mathcal{S} + \mathcal{S}, \quad \text{where} \quad \mathcal{S}\equiv 2\kappa a \cdot a^{\dagger} \equiv J \cdot J^{\dagger} \quad \text{with} \quad J=\sqrt{2\kappa}a, \quad \text{and} \quad (\mathcal{L}-\mathcal{S})=-\kappa(a^{\dagger}a\cdot - \cdot a^{\dagger}a).
\end{equation}
A similar expression holds for $\mathcal{L}_{\rm kappa}^{\rm JCD}$, since the dissipation channel is the same. Collapses occur with a conditional probability density ${\rm tr}_S[\mathcal{S}\rho_{\rm REC}(t)]$, where $\rho_{\rm REC}(t)=\overline{\rho}_{\rm REC}(t)/{\rm tr_S[ \overline{\rho}_{\rm REC}(t)}]$, while in between collapses the conditional state evolves as $d\,\overline{\rho}_{\rm REC}(t)/dt=(\mathcal{L}-\mathcal{S})\overline{\rho}_{\rm REC}(t)$. 

When the conditioned density operator factorizes as a pure state and satisfies a non-unitary Schr\"{o}dinger equation between the collapses, the following Monte Carlo algorithm is used to advance the conditional ket and generate trajectories, entailing four steps:

\noindent {\bf 1.} Compute the conditional mean photon number 
\begin{equation}
\langle n (t) \rangle_{\rm REC}\equiv\langle (a^{\dagger}a)(t)\rangle_{\rm REC}=\frac{\langle \overline{\psi}_{\rm REC}(t)|(a^{\dagger}a)|\overline{\psi}_{\rm REC}(t)\rangle}{\langle \overline{\psi}_{\rm REC}(t)|\overline{\psi}_{\rm REC}(t)\rangle}.
\end{equation}
The quantity $p_{\rm REC}(t)=2\kappa \langle n (t) \rangle_{\rm REC} \Delta t$ is the conditioned probability for a collapse in the interval $(t,t+\Delta t]$. 

\noindent{\bf 2.} Generate a random number $r$ uniformly distributed along the unit line.

\noindent{\bf 3.} If $p_{\rm REC}(t)>r$, advance the conditional state by executing a quantum jump (photon emission):
\begin{equation}
|\overline{\psi}_{\rm REC}(t)\rangle \to |\overline{\psi}_{\rm REC}(t+\Delta t)\rangle=J |\overline{\psi}_{\rm REC}(t)\rangle.
\end{equation}

\noindent{\bf 4.} If $p_{\rm REC}(t)<r$, then advance the conditional state under the action of the ``free'' propagation non-Hermitian Hamiltonian:
\begin{equation}
|\overline{\psi}_{\rm REC}(t)\rangle \to |\overline{\psi}_{\rm REC}(t+\Delta t)\rangle=\exp\left(\frac{1}{i\hbar} H \Delta t \right)|\overline{\psi}_{\rm REC}(t)\rangle.
\end{equation}
For an ensemble of states $|\psi^{(k)}_{\rm REC}(t)\rangle$, $k=1,2,\ldots N$, generated in this way, the expansion of $\rho$ as a sum over records is approximated by
\begin{equation}
\rho(t)=\frac{1}{N}\sum_{k=1}^{N}|\psi^{(k)}_{\rm REC}(t)\rangle \langle \psi^{(k)}_{\rm REC}(t)|=\frac{1}{N}\sum_{k=1}^{N}\rho^{(k)}_{\rm REC}(t).
\end{equation}

A quantum jump at a time $t_k$ in the interval $(t,t+\Delta t]$ which increases the conditional photon number, $\langle (a^{\dagger}a)(t_k^{+})\rangle_{\rm REC}>\langle (a^{\dagger}a)(t_k^{-})\rangle_{\rm REC}$, is termed {\it upward jump} (or {\it upward collapse}). Otherwise, when $\langle (a^{\dagger}a)(t_k^{+})\rangle_{\rm REC}<\langle (a^{\dagger}a)(t_k^{-})\rangle_{\rm REC}$, we speak of a downward jump. This means that the conditional collapse probability increases as this type of jump adds information to the memory; note that the state is conditioned in principle on everything that has taken place along the trajectory in the past. Upward jumps substantiate bunching in a thermal state: the observation of one photon emission means that another one is very likely -- at twice the average rate -- immediately following the first.  

The non-Hermitian Hamiltonian takes the form $H=H_{\rm JCD}-i\hbar \kappa a^{\dagger}a$ to generate the trajectories of complex-amplitude bistability in the JC model, or simply $H=-i\hbar \kappa a^{\dagger}a$ when we deal with the decay of a coherent-state superposition in an empty cavity. When spontaneous emission at rate $\gamma$ is included, the total jump probability $p_{\rm REC}(t)=[2\kappa \langle (a^{\dagger}a)(t)\rangle_{\rm REC} + \gamma \langle (\sigma_{+}\sigma_{-})(t)\rangle_{\rm REC}]\Delta t$ is calculated in Step 1, and then two random numbers $r_1, r_2$ on the unit line are generated in Step 2. Step 3 is divided into two sub-steps: one to decide whether a jump occurs or not (Step 3a with $r_1$) and another one to determine what kind of jump occurred---either a cavity or a spontaneous emission (Step 3b with $r_2$). In Step 4, the non-Hermitian Hamiltonian of the dissipative JC model is instead: $H=H_{\rm JCD}-i\hbar \kappa a^{\dagger}a-i\hbar(\gamma/2) \sigma_{+}\sigma_{-}$.

The ket in Step 4 of the Monte Carlo algorithm is numerically advanced via the implementation of a Runge-Kutta method of 4$^{\rm th}$ order with the $3/8$ rule, using a truncated basis $\{|n\rangle|-\rangle, |n\rangle|+\rangle; n=0, 1,\ldots L_{\rm max}\}$, where $|n\rangle$ denotes the Fock states, and $|+\rangle$, $|-\rangle$ are the upper and lower states of the atom, respectively. For example, the results reported for single realizations in the main text have been generated with $L_{\rm max}=25$ in Fig.~\ref{fig:FIG2}, and $L_{\rm max}=60$ in Figs.~\ref{fig:FIG3} and~\ref{fig:FIG12}. Convergence has been checked against the time step $\kappa\Delta t=0.002 \ll \kappa/g$ and $L_{\rm max}$. The initial state is taken pure and is everywhere set to the ground JC eigenstate $|0\rangle|-\rangle$. On the other hand, ensemble-averaged steady-state quantities, for example the functions $Q_{\rm ss}(x+iy)$ depicted in the figures of the main text, are obtained via a numerical diagonalization and exponential series expansion of the Liouvillian in \textsc{Matlab}'s {\it Quantum Optics Toolbox}~\cite{Tan1999}. The Fock-state basis is truncated at $L_{\rm max}=25$ in Fig.~\ref{fig:FIG2}, and $L_{\rm max}=70$ in Figs.~\ref{fig:FIG3} and~\ref{fig:FIG12} (in the relevant frames).

In the truncated Fock-state basis, the matrix elements of the conditional cavity-field state along an individual record are 
\begin{equation}
\langle m|\rho_{\rm cav, REC}(t)|n \rangle = \langle m (\langle +|\rho_{\rm REC}(t)|+\rangle + \langle -|\rho_{\rm REC}(t)|-\rangle) | n \rangle.
\end{equation}
These elements are used to calculate the $Q$-function of the conditional field state as
\begin{equation}
Q_{\rm REC}(x+iy;t)=\langle \alpha |\rho_{\rm cav, REC}| \alpha\rangle=\frac{1}{\pi} e^{-|\alpha|^2}\sum_{m,n}\frac{\alpha^{*m}\alpha^n}{\sqrt{n!m!}}\langle m|\rho_{\rm cav, REC}(t)|n \rangle,
\end{equation}
where $\alpha=x+iy$ is the phase-space variable, together with $\alpha^{*}=x-iy$. 

\subsection{Sources of incoherence: thermally excited bath and imperfect detection}
\label{subsec:sourcesinc}

Let us now consider the decay of a coherent-state superposition in an empty cavity weakly coupled to a reservoir in a thermal state with mean excitation $\overline{n}$. Since the excitation is incoherent, we are unable the factorize the conditioned density operator as a pure state. Between collapses, the conditional state satisfies:
\begin{equation}
\frac{d\,\overline{\rho}_{\rm REC}}{dt}=(\mathcal{L}-\mathcal{S})\overline{\rho}_{\rm REC}=-\kappa(a^{\dagger}a \overline{\rho}_{\rm REC} + \overline{\rho}_{\rm REC} a^{\dagger}a)+2\kappa \overline{n}(a\overline{\rho}_{\rm REC} a^{\dagger} + a^{\dagger} \overline{\rho}_{\rm REC} a - a^{\dagger}a \overline{\rho}_{\rm REC}-\overline{\rho}_{\rm REC}a^{\dagger}a). 
\end{equation} 
The term proportional to $\overline{n}$ prevents us from using a pure state between collapses. 

In quantum optical experiments, the detectors intercepting the output beam register photon ``clicks'' with a limited efficiency $\eta <1$. This will modify the collapse probability in the interval $(t, t+\Delta t]$ as $p_{\rm REC}(t)=\eta{\rm tr}_S[\mathcal{S}\rho_{\rm REC}(t)]\Delta t=\eta (2\kappa){\rm tr}_S[\rho_{\rm REC}(t)a^{\dagger}a]$, while the continuous evolution will be governed by the superoperator $\mathcal{L}-\eta\mathcal{S}$. 
 
Adding the two effects, the non-Hermitian evolution corresponding to Step 4 is replaced by the following matrix-element equation for the un-normalized conditional state $\overline{\rho}_{\rm REC}(t)$ in a truncated Fock-state basis:
\begin{equation}\label{eq:MatrixElements}
\begin{aligned}
\left(\frac{d\, \overline{\rho}_{\rm REC}}{dt}\right)_{mn}=&-\kappa\Big\{[(\overline{n}+1)(m+n)+\overline{n}(m+n+2)](\overline{\rho}_{\rm REC})_{mn}-2(\overline{n}+1)\sqrt{(m+1)(n+1)} (\overline{\rho}_{\rm REC})_{m+1,n+1}\\
&-2\overline{n}\sqrt{mn} (\overline{\rho}_{\rm REC})_{m-1,n-1} -2(1-\eta)\sqrt{(m+1)(n+1)} (\overline{\rho}_{\rm REC})_{m+1,n+1} \Big\}.
\end{aligned}
\end{equation}  
The conditional jump probability to be compared against the random number $r$ is
\begin{equation}
p_{\rm REC}(t)=\eta (2\kappa){\rm tr}_S[\rho_{\rm REC}(t)a^{\dagger}a]\Delta t=\eta (2\kappa)\sum_{n=0}^{L_{\rm max}} n (\rho_{\rm REC}(t))_{nn}\,\Delta t,
\end{equation}
where $L_{\rm max}$ is the truncation level of the Fock-state basis. The matrix elements of the conditional matrix corresponding to Eq.~\eqref{eq:MatrixElements} are numerically advanced via a Runge--Kutta--Fehlberg method of $5^{\rm th}$ order. 

\section{Damped coherent states and coherent-state superpositions}

\subsection{A single damped coherent state: the Poisson distribution operationally derived}
\label{subsubsec:dampedA}

In this section, we closely follow the treatment of~\cite{Carmichael2013Ch4}. We first consider a damped coherent state. Working in a frame rotating with the resonant mode frequency, we aim to solve the ME~\eqref{eq:MEd}:
\begin{equation}
\frac{d\rho}{dt}=\mathcal{L}\rho=\kappa(2a\rho a^{\dagger}-a^{\dagger}a\rho-\rho a^{\dagger}a)
\end{equation}
describing the decay of the cavity from an initially assigned coherent state $|\alpha(0)\rangle$. We substitute the {\it ansatz} $\rho(t)=|\alpha(t)\rangle\langle \alpha(t)|$ which preserves coherence throughout the decay. Hence, the state $|\alpha(t)\rangle$ must obey the equation of motion
\begin{equation}\label{eq:eqS}
\frac{d|\alpha(t)\rangle}{dt}=\kappa(\alpha^{*}(t)\alpha(t)-\alpha(t) a^{\dagger})|\alpha(t)\rangle.
\end{equation}
Now, any coherent state can be generated from the vacuum as
\begin{equation}
|\alpha(t) \rangle=\exp(-|\alpha(t)|^2/2)\exp(\alpha(t) a^{\dagger})|0\rangle.
\end{equation}
Differentiating with respect to time we obtain:
\begin{equation}\label{eq:eqG}
\frac{d|\alpha(t)\rangle}{dt}=\left[-\frac{1}{2}\left(\frac{d\alpha^{*}(t)}{dt}\alpha(t) + \alpha^{*}(t)\frac{d\alpha(t)}{dt}\right) + \frac{d\alpha(t)}{dt}a^{\dagger}\right]|\alpha(t)\rangle.
\end{equation}
Equations~\eqref{eq:eqS} and~\eqref{eq:eqG} are consistent if $d\alpha(t)/dt=-\kappa \alpha(t)$, in which case the {\it ansatz} works, producing the following solution the the ME~\eqref{eq:MEd}:
\begin{equation}
\rho(t)=|\alpha(0)e^{-\kappa t}\rangle \langle \alpha(0)e^{-\kappa t}|. 
\end{equation}
To generate photoelectron counting records, we note that provided the initial state is pure, the conditioned state also factorizes as a pure state in the unraveling of the ME~\eqref{eq:MEd}. As we saw in Appendix~\ref{sec:MC}, the Liouvillian $\mathcal{L}$ can be split with 
\begin{equation}
\mathcal{S}=2\kappa\,a\cdot a^{\dagger}, \quad \mathcal{L}-\mathcal{S}=\frac{1}{i\hbar}(H\cdot-\cdot H^{\dagger}), \quad H=-i\hbar \kappa a^{\dagger}a.
\end{equation}
The full Dyson expansion
\begin{equation}
\rho(t)=\sum_{n=0}^{\infty}\int_{0}^{t}dt_n \int_{0}^{t_n}dt_{n-1}\cdots\int_{0}^{t_2}dt_1\,e^{(\mathcal{L}-\mathcal{S})(t-t_n)}\mathcal{S}e^{(\mathcal{L}-\mathcal{S})(t_n-t_{n-1})}\mathcal{S}\cdots \mathcal{S}e^{(\mathcal{L}-\mathcal{S})t_1}\rho(0),
\end{equation}
with the assumption of a pure initial state, can be recast to the form
\begin{equation}
\rho(t)=\sum_{n=0}^{\infty}\int_{0}^{t}dt_n \int_{0}^{t_n}dt_{n-1}\cdots\int_{0}^{t_2}dt_1\,P_{\rm REC}(t)|\psi_{\rm REC}(t)\rangle \langle \psi_{\rm REC}(t)|,
\end{equation}
with record probability
\begin{equation}\label{eq:ProbRec}
P_{\rm REC}(t)dt_1\ldots dt_n=\langle \overline{\psi}_{\rm REC}(t)|\overline{\psi}_{\rm REC}(t)\rangle\,dt_1\ldots dt_n
\end{equation}
and conditional states
\begin{equation}
|\psi_{\rm REC}(t)\rangle=\frac{|\overline{\psi}_{\rm REC}(t)\rangle}{\sqrt{\langle\overline{\psi}_{\rm REC}(t)|\overline{\psi}_{\rm REC}(t)\rangle}},
\end{equation}
where the un-normalized conditional state assumes the form
\begin{equation}\label{eq:PsiUn}
|\overline{\psi}_{\rm REC}(t)\rangle=\exp\left[\frac{1}{i\hbar}H(t-t_n)\right]J \exp\left[\frac{1}{i\hbar}H(t_n-t_{n-1})\right]J \cdots J \exp\left[\frac{1}{i\hbar}H t_1\right]|\psi(0)\rangle,
\end{equation}
with $J=\sqrt{2\kappa}\,a$ the jump operator defined as $\mathcal{S}=J \cdot J^{\dagger}$. The quantity $P_{\rm REC}(t)$ so defined is the exclusive probability density for counting precisely $n$ photons in the output field at times $t_1, t_2, \ldots, t_n$ and no photon at any other time in the interval from $0$ to $t$. 

Let us now follow the procedure suggested by Eq.~\eqref{eq:PsiUn} to generate a photoelectron counting record for a damped cavity resonator governed by the ME~\eqref{eq:MEd}. The initial state is set to $|\overline{\psi}_{\rm REC}(0)\rangle=|\alpha\rangle$, which --on the basis of the solution to the ME -- suggests to adopt the {\it ansatz}
\begin{equation}\label{eq:ansatz2}
|\overline{\psi}_{\rm REC}(t)\rangle=A(t)|\alpha(t)\rangle, \quad \text{with}\quad |\alpha(t)\rangle=\exp[\alpha(t)a^{\dagger}-\alpha^{*}(t)a]|0\rangle.
\end{equation}
From Eq.~\eqref{eq:ProbRec}, we find that the record probability density is given by the norm $\langle \overline{\psi}_{\rm REC}(t)|\overline{\psi}_{\rm REC}(t)\rangle=|A(t)|^2$. The solution pertains not only to a ME but also to a photon counting problem. 

Let us consider a record of $n$ counts up to time $t$. The conditioned state evolves continuously between quantum jumps which take place at the ordered count times $t_1, t_2,\ldots t_n$. A photon emission at time $t_k$ preserves the {\it ansatz} while modifying the norm in the following way:
\begin{equation}
A(t_k) \to \sqrt{2\kappa}\,\alpha(t_k)A(t_k).
\end{equation}
Now, the evolution between jumps is dictated by a Schr\"{o}dinger equation with the non-Hermitian Hamiltonian $H=-i\hbar \kappa a^{\dagger}a$. The resulting equation of motion
\begin{equation}
\frac{d |\overline{\psi}_{\rm REC}(t)\rangle}{dt}=-\kappa(a^{\dagger}a)|\overline{\psi}_{\rm REC}(t)\rangle
\end{equation}
also preserves the {\it ansatz}~\eqref{eq:ansatz2}, provided the amplitudes $\alpha(t)$ and $A(t)$ satisfy the equations
\begin{equation}
\frac{d\alpha}{dt}=-\kappa \alpha, \quad \text{and} \quad \frac{1}{A}\frac{dA}{dt}=-\frac{d\alpha^{*}}{dt}\alpha=-\kappa|\alpha|^2.
\end{equation}
The above equations have solutions
\begin{equation}
\alpha(t)=\alpha \exp(-\kappa t), \quad \text{and} \quad A(t)=A(t_k)\exp\left[-\tfrac{1}{2}|\alpha|^2 (e^{-2\kappa t_k}-e^{-2\kappa t})\right], \quad t_k\leq t<t_{k+1}.
\end{equation}
Putting the pieces together, we can produce an analytical expression for the conditional state~\eqref{eq:PsiUn},
\begin{equation}\label{eq:psicona}
|\overline{\psi}_{\rm REC}(t)\rangle=(\sqrt{2\kappa}\,\alpha\,e^{-\kappa t_n})\ldots (\sqrt{2\kappa}\,\alpha\,e^{-\kappa t_1})\,\exp\left[-\tfrac{1}{2}|\alpha|^2 (1-e^{-2\kappa t})\right]\,|\alpha e^{-\kappa t}\rangle,
\end{equation}
from which we calculate the conditional probability density
\begin{equation}
\langle \overline{\psi}_{\rm REC}(t)| \overline{\psi}_{\rm REC}(t) \rangle = (2\kappa\, |\alpha|^2\, e^{-2\kappa t_n})\ldots (2\kappa\, |\alpha|^2\, e^{-2\kappa t_1}) \exp[-|\alpha|^2(1-e^{-2\kappa t})].
\end{equation}
This result can be construed on the basis of photoelectron counting for a classical field with decaying intensity $I(t)=2\kappa |\alpha|^2 \exp(-2\kappa t)$ in photon flux units. The exponential term $\exp[-|\alpha|^2(1-e^{-2\kappa t})]$ is the probability for no photon counts over an interval of length $t$. This instance can be appreciated by the expression of the probability for registering $n$ photon ```clicks'' in an interval $[0,t)$ for a time-varying light intensity $I(t)$,
\begin{equation}\label{eq:Pn0t}
P_{n;[0,t)}=\frac{\left(\int_{0}^{t} I(t^{\prime})dt^{\prime}\right)^n}{n!}\exp\left[-\left(\int_{0}^{t} I(t^{\prime})dt^{\prime}\right)\right].
\end{equation} 
Applying of the above formula for $n=0$ we obtain $\exp[-|\alpha|^2(1-e^{-2\kappa t})]$ as the probability for a sequence of null-measurement results. Next, $2\kappa |\alpha^2| e^{-2\kappa t_{k}}=\langle \alpha e^{-\kappa t}|J^{\dagger}J| \alpha e^{-\kappa t} \rangle$ (Step 1 of the Monte Carlo algorithm) is the probability density for a photon count at time $t_k$. Integrating the record probability density over all  possible count times after using
\begin{equation}
(2\kappa |\alpha|^2)^n \int_{0}^{t}dt_n \int_{0}^{t_n}dt_{n-1}\cdots\int_{0}^{t_2}dt_1 e^{-2\kappa t_n}\ldots e^{-2\kappa t_1}=\frac{\left[2\kappa|\alpha|^2 \int_{0}^{t}e^{-2\kappa t^{\prime}}dt^{\prime}\right]^n}{n!},
\end{equation} 
takes us to the Poisson distribution~\eqref{eq:Pn0t}:
\begin{equation}
P_{n;[0,t)}=\frac{[|\alpha|^2 (1-e^{-2\kappa t})]^n}{n!}\exp[-|\alpha|^2 (1-e^{-2\kappa t})].
\end{equation} 

\subsection{Damped coherent-state superposition and localization time}
\label{subsubsec:dampedcs}

Armed with these observations, let us now consider the damped coherent state superposition which we extensively discuss in this work. We take the initial state
\begin{equation}\label{eq:inputState}
|\overline{\psi}_{\rm REC}(0)\rangle=\frac{1}{\sqrt{2}}\frac{|\alpha_1\rangle + |\alpha_2\rangle}{1+{\rm Re}\langle \alpha_1|\alpha_2\rangle},
\end{equation}
where $\alpha_1$ and $\alpha_2$ are any two complex numbers. The Schr\"{o}dinger equation dictating the evolution of the un-normalized conditional state is linear, whence $|\overline{\psi}_{\rm REC}(t) \rangle$ is a sum of two terms, each having the form of Eq.~\eqref{eq:psicona},
\begin{equation}
|\overline{\psi}_{\rm REC}(t) \rangle=|\overline{\psi}^{(\alpha_1)}_{\rm REC}(t) \rangle + |\overline{\psi}^{(\alpha_2)}_{\rm REC}(t) \rangle,
\end{equation}
where
\begin{equation}
|\overline{\psi}^{(\beta)}_{\rm REC}(t) \rangle=(\sqrt{2\kappa}\,\beta\,e^{-\kappa t_n})\ldots (\sqrt{2\kappa}\,\beta\,e^{-\kappa t_1})\,\exp\left[-\tfrac{1}{2}|\beta|^2 (1-e^{-2\kappa t})\right]\,|\beta e^{-\kappa t}\rangle, \quad \beta=\alpha_1, \alpha_2.
\end{equation}
In the general case where $|\alpha_1|$ and $|\alpha_2|$ are different, the two components $|\overline{\psi}^{(\alpha_1)}_{\rm REC}(t) \rangle$ and $|\overline{\psi}^{(\alpha_2)}_{\rm REC}(t) \rangle
$ are in competition with each other to dominate the sum, since one component increases in norm relative to the other. The sequence of count times in a trajectory determines the outcome in the following fashion: a small number of photon ``clicks'' causes the component corresponding to the smaller of $|\alpha_1|$ and $|\alpha_2|$ to dominate, since the continuous evolution in between the jumps, $A_{\beta}(t)=A_{\beta}(t_k) \exp[-\tfrac{1}{2}|\beta|^2(e^{-2\kappa t_k}-e^{-2\kappa t})]$, $t_k\leq t <t_{k+1}$, favours a larger $|\beta|^2$. On the other hand, a big number of counts promotes the component corresponding to the larger of $|\alpha_1|$ and $|\alpha_2|$, since the jump evolution $A_{\beta}(t_k) \to \sqrt{2\kappa}\,\beta(t_k)\, A_{\beta}(t_k)$ favours a larger $|\beta|^2$. In each realization, the state localizes into the one or the other component of the superposition, while the average over trajectories produces a statistical mixture as the steady state of ME~\eqref{eq:MEd}. For an individual record, we can say that decoherence manifests itself my making a choice between alternative ``meter'' states in a quantum measurement. 

The quantum jumps of JC amplitude bistability we are concerned with are modelled by the above-mentioned localization in the course of a null-measurement (no ``click'') evolution producing a conditional state:
\begin{equation}
|\overline{\psi}_{\rm REC}(t) \rangle=\exp\left[-\tfrac{1}{2}|\alpha_1|^2 (1-e^{-2\kappa t})\right]\,|\alpha_1 e^{-\kappa t}\rangle + \exp\left[-\tfrac{1}{2}|\alpha_2|^2 (1-e^{-2\kappa t})\right]\,|\alpha_2 e^{-\kappa t}\rangle.
\end{equation}  
and a conditional photon number
\begin{equation}\label{eq:NrecNull}
\begin{aligned}
\langle n(t)\rangle_{\rm REC, NULL}&=\frac{\langle \overline{\psi}_{\rm REC, NULL} (t)|a^{\dagger}a|\overline{\psi}_{\rm REC, NULL} (t)\rangle}{\langle \overline{\psi}_{\rm REC, NULL} (t)|\overline{\psi}_{\rm REC, NULL} (t)\rangle}\\
&=\frac{|\alpha_1(t)|^2\lambda(|\alpha_1|^2;t) + |\alpha_2(t)|^2\lambda(|\alpha_2|^2;t)+2\lambda(|\alpha_1|^2/2;t)\lambda(|\alpha_2|^2/2;t)\,{\rm Re}(\alpha_1^{*}(t)\alpha_2(t)\langle\alpha_1(t)|\alpha_2(t)\rangle)}{\lambda(|\alpha_1|^2;t)+\lambda(|\alpha_2|^2;t)+2\lambda(|\alpha_1|^2/2;t)\lambda(|\alpha_2|^2/2;t)\,{\rm Re}(\langle\alpha_1(t)|\alpha_2(t)\rangle)},
\end{aligned}
\end{equation}
where we have defined $\lambda(x;t)\equiv \exp[-x(1-e^{-2\kappa t})]$ to denote the factor arising from the null-measurement evolution of each component. The decaying amplitudes of the two components are $\alpha_1(t)=\alpha_1 e^{-\kappa t}$, $\alpha_2(t)=\alpha_2 e^{-\kappa t}$.  

For a large separation between the two coherent states, $|\alpha_1-\alpha_2|^2$, recalling that
\begin{equation}
\langle \alpha_1 | \alpha_2 \rangle = e^{-\frac{1}{2}|\alpha_1|^2} e^{-\frac{1}{2}|\alpha_2|^2}e^{\alpha_2^{*}\alpha_1}, \quad \text{whence} \quad |\langle \alpha_1 | \alpha_2\rangle|^2=e^{-|\alpha_1-\alpha_2|^2},
\end{equation}
we can approximate Eq.~\eqref{eq:NrecNull} by
\begin{equation}\label{eq:NullRecApp}
\langle n(t) \rangle_{\rm REC, NULL} \approx \frac{|\alpha_1(t)|^2\lambda(|\alpha_1|^2;t) + |\alpha_2(t)|^2\lambda(|\alpha_2|^2;t)}{\lambda(|\alpha_1|^2;t)+\lambda(|\alpha_2|^2;t)},
\end{equation}
which can be employed to estimate the duration of the quantum jump from the bright state (amplitude $\alpha_1$) to the unstable state (amplitude $\alpha_2$). 

To obtain a lower bound for that duration $\kappa (t_{\rm end}-0)=\kappa(t^{\prime}-t^{\prime\prime})/2=\kappa \Delta t_{\rm end} \ll 1$ (see the prescription of Sec.~\ref{subsec:cohlocJC} in the main text), we formulate a semiclassical argument. Instead of solving $n_{\rm REC, NULL}(t)=|\alpha_2|^2$, we formulate the alternative equation
\begin{equation}\label{eq:durationapp2}
|\alpha_1|^2\exp(-2\kappa |\alpha_1|^2 t) + |\alpha_2|^2\exp(-2\kappa |\alpha_2|^2 t)=|\alpha_2|(|\alpha_2|-1)\exp(-2\kappa |\alpha_1|^2 t)+ |\alpha_2|(|\alpha_2|+1) \exp(-2\kappa |\alpha_2|^2 t),
\end{equation} 
suggesting that the photon number in the unstable state fluctuates within two standard deviations of the Poisson distribution, $2\sqrt{|\alpha_2|^2}=2|\alpha_2|$, in the course of the switch. A standard deviation is subtracted (added) when the bright (unstable) state dominates. The spiralling peak of the unstable state and its splitting into a complex of states observed in the conditional phase-space distributions motivate this modification. In producing Eq.~\eqref{eq:durationapp2}, we have set $1-\exp(-2\kappa t) \approx 2\kappa t$ in the exponential of the $\lambda$-terms, but $e^{-\kappa t} \approx 1$ in the decaying coherent-state amplitudes. The result has
\begin{equation}\label{eq:durationLoc}
\Delta t_{\rm end,\,min}=-\frac{1}{2\kappa (|\alpha_1|^2-|\alpha_2|^2)}\ln\left(\displaystyle\frac{|\alpha_2|}{|\alpha_1|^2-|\alpha_2|(|\alpha_2|-1)}\right).
\end{equation}
Calculating the standard deviation counteracts the approximation effected in the $\lambda$-terms. Thus, in the ``thermodynamic limit'' $n_{\rm scale}=[g/(2\kappa)]^2 \to \infty$, where the two states become increasingly separated, we approach the graphical intersection of $|\alpha_1(t)|^2\lambda(|\alpha_1|^2;t) + |\alpha_2(t)|^2\lambda(|\alpha_2|^2;t)$ and $|\alpha_2|^2 [\lambda(|\alpha_1|^2;t)+\lambda(|\alpha_2|^2;t)]$ proposed in Sec. III B of the main text as a means to determine the localization time. 

We finally note that inversion about the localization mid-point is conceptually different from time reversal $t\to -t$ in $n_{\rm REC, NULL}(t)$ about the time origin set to coincide with the midpoint. The null-measurement record is irreversible, since it is governed by a non-Hermitian Hamiltonian. Nevertheless, the relative difference between the conditioned photon number in the two processes, in the interval required for the transition from $|\alpha_2|^2$ to $|\alpha_1|^2$, decreases when $|\alpha_1|\gg |\alpha_2|$---this is the case for the high-amplitude nonlinearity in the  ``thermodynamic limit'' $n_{\rm scale}=g^2/(4\kappa^2) \to \infty$, where $\kappa \Delta t_{\rm end} \to 0$. 

\vspace{3mm}

{\bf Data availability}: The data that support the findings reported herein are openly available in {\it figshare} at the set with this \href{http://doi.org/10.17045/sthlmuni.30122545}{DOI}.

\acknowledgements

Work supported by the Swedish Research Council (VR) and the Severo Ochoa Centre of Excellence.

\twocolumngrid

\bibliography{bibliography_LV}
\end{document}